\documentclass[a4paper,12pt]{article}

\usepackage{amsmath,amssymb,amsfonts}
\usepackage{epsfig,cancel,cite}
\usepackage{dsfont}

\newlength{\xtrawidth}
\newlength{\xtraheight}
\newcommand{\calk}{{\mathcal K}}
\newcommand{\nn}{\nonumber}

\def\fnote#1#2{\begingroup\def\thefootnote{#1}\footnote{#2}
     \addtocounter{footnote}{-1}\endgroup}

\newcommand{\setall}{\setcounter{equation}{0}}

\begin{document}

\title{{\LARGE G-structures and Domain Walls in Heterotic Theories}}
\author{Andre Lukas and Cyril Matti}
\date{}
\maketitle
\begin{center}
{\small {\it Rudolf Peierls Center for Theoretical Physics, Oxford University,\\
$~~~~~$ 1 Keble Road, Oxford, OX1 3NP, U.K.}\\
\fnote{}{lukas@physics.ox.ac.uk}
\fnote{}{c.matti1@physics.ox.ac.uk}
}
\end{center}

\abstract{We consider heterotic string solutions based on a warped product of a four-dimensional domain wall and a six-dimensional internal manifold, preserving two supercharges. The constraints on the internal manifolds with $SU(3)$ structure are derived. They are found to be generalized half-flat manifolds with a particular pattern of torsion classes and they include half-flat manifolds and Strominger's complex non-Kahler manifolds as special cases. We also verify that previous heterotic compactifications on half-flat mirror manifolds are based on this class of solutions.}

\newpage

\tableofcontents

\section{Introduction}
\setall

Compactifications of the heterotic string on Calabi-Yau manifolds has been a successful avenue towards string model building since the early days of string theory~\cite{Candelas:1985en} and with recent progress~\cite{Braun:2005ux,Bouchard:2005ag,Anderson:2009mh} models are edging closer and closer to a realistic standard model from string theory. However, part of any program aiming at realistic string models must be the stabilization of moduli and this is where heterotic compactifications encounter problems. In type II models a combination of NS and RR flux allows one, at least in principle, to stabilize all complex structure moduli and the dilaton while, thanks to the no-scale structure, keeping the theory in a Minkowski vacuum~\cite{Giddings:2001yu}. In heterotic compactifications, on the other hand, only NS flux is available. This stabilizes the complex structure moduli only and due to the absence of the no-scale property the theory does not remain in a Minkowski vacuum. Instead, the simplest four-dimensional vacuum solution in the presence of NS flux can be expected to be a domain wall. Another problematic feature is that the heterotic flux superpotential, unlike its IIB counterpart, does not allow itself to be tuned to small values by a careful choice of the flux integers. All these features mean that it will be difficult at best to achieve a scale separation between the string and the flux scale in heterotic Calabi-Yau models with flux. 

The previous discussion suggests that heterotic models with stable moduli may require compactifications on more general manifolds with $SU(3)$ structure, where some of the ``missing" RR flux is replaced by the intrinsic torsion of the manifold. Studying such more general backgrounds for heterotic compactifications is the main purpose of the present paper. One such class of compactifications has been identified early on by Strominger~\cite{Strominger:1986uh}. To obtain this class a maximally symmetric four-dimensional space and four preserved supercharges have been assumed. In this case, it turns out that the associated internal six-dimensional manifolds have $SU(3)$ structure and are complex but, in general, no longer Kahler. In the present paper, we will generalize this discussion by relaxing both initial assumptions. We will allow the four-dimensional space to deviate from maximal symmetry, more specifically, we will allow it to be a domain wall, and we will only require two preserved supercharges for the 10-dimensional solution. 

Why are we interested in backgrounds which violate the conventional requirement of a four-dimensional maximally symmetric space?
The simple answer is that string compactifications with flux often do not allow for a maximally symmetric four-dimensional space, unless special conditions such as the no-scale structure are realized. Frequently, a flux superpotential in four dimensions leads to a runaway potential for some of the not-yet stabilized moduli and the simplest solution consistent with this feature is a four-dimensional domain wall. 
This happens for heterotic Calabi-Yau compactifications with flux but also for more general heterotic compactifications on half-flat manifolds as studied in Refs.~\cite{Gurrieri:2004dt, Gurrieri:2007jg, deCarlos:2005kh, Gurrieri:2005af}.  A stabilization of all moduli presumably requires some non-perturbative effects which are not normally included when studying 10-dimensional solutions of string theory. Phenomenologically, one should require a four-dimensional maximally symmetric space after all relevant effects, including non-perturbative ones, have been included. When studying 10-dimensional perturbative string solutions we should, therefore, allow for more general four-dimensional spaces, keeping in mind the possibility of a non-perturbative ``lift" to a maximally symmetric four-dimensional space.

With this motivation in mind, we will study 10-dimensional solutions of the heterotic string which consist of a warped product of a six-dimensional internal space and a four-dimensional domain wall and preserve two supercharges (that is, they are half BPS from a four-dimensional $N=1$ point of view). There are two main questions we would like to answer in this context. First, what are the allowed internal six-dimensional spaces in such a setting? This question will be answered using the G-structure formalism~\cite{Hitchin,ChiossiSalamon} (for a review see \cite{Grana:2005jc}) applied to the heterotic case~\cite{LopesCardoso:2002hd}, for the groups $G_2$ and $SU(3)$,  and this leads to a significant generalization of the class of manifolds found by Strominger. Secondly, we would like to show the consistency of certain heterotic compactifications on half-flat mirror manifolds \cite{Gurrieri:2004dt, Gurrieri:2007jg, deCarlos:2005kh} which has been carried out in the absence of a full 10-dimensional solution. This will be done by verifying that such half-flat mirror manifolds are allowed internal manifolds within our generalized setting
and that the domain wall solutions in the associated four-dimensional $N=1$ supergravity theories do indeed lift up to the correct 10-dimensional solutions. A general classification of solutions to the heterotic string which preserve some supersymmetry has been carried out in Ref.~\cite{Gran:2005wf}, using spinorial methods. The solutions considered in the present paper fit into this classification and correspond to cases with two supercharges and $G_2$ stability group. 

In this paper, we will work to zeroth order in $\alpha'$, that is, we will not consider gauge fields explicitly although the standard embedding should provide at least one way of completing our models to include gauge fields.

The plan of the paper is as follows. In the next section, we set the scene by reviewing some general properties of the heterotic string and by defining our solution Ansatz. In Section 3 we focus on the case of vanishing flux and constant dilaton, as a warm-up. We derive the structure of the 10-dimensional solutions in this case and show that they can be matched up with four-dimensional domain wall solutions in the associated compactified theories. We repeat this discussion but in full generality including non-vanishing flux and a non-constant dilaton in Section 4. Finally, in Section 5 we introduce a class of Calabi-Yau domain wall solutions with flux. We conclude in Section 6. Three technical appendices set out our conventions, review $G$-structures and their associated torsion classes for the groups $G_2$ and $SU(3)$ and collect some relevant formulae for Calabi-Yau moduli spaces.

\section{Ten-dimensional theory and solution Ansatz}\label{Heterotic KSE}

To set the scene, we briefly review  the 10-dimensional effective action of the heterotic string and its associated Killing spinor equations (see, for example, Refs.~\cite{Green:1987mn,polchinski} for details). Then we discuss our solution Ansatz.

\subsection{Action and Killing spinor equation}
The bosonic spectrum of this effective theory consists of the 10-dimensional metric $\hat{G}_{MN}$, the dilaton $\hat\phi$ and the NS-NS rank two anti-symmetric tensor field $\hat{B}=\frac{1}{2}\hat B_{MN}dx^M\wedge dx^N$ with field strength
\begin{equation}
 \hat H=d\hat B\; . \label{BI}
\end{equation}
Here, we use indices $M,N,\ldots = 0,1,\ldots ,9$ to label the 10-dimensional space-time coordinates $x^M$. (For a summary of our index conventions see Appendix~\ref{Conventions}.) In addition, we have the gauge field $A_M$ with associated gauge group $SO(32)$ or $E_8\times E_8$ and field strength $F_{MN}$. To lowest order in $\alpha '$, the bosonic part of the string frame action is given by
\begin{equation}\label{action10d}
	S^S_{0,{\rm bosonic}}=-\frac{1}{2\kappa^2_{10}}\int_{M_{10}} e^{-2\hat\phi}\left[\hat{R}*\textbf{1}-4d\hat\phi\wedge *d\hat\phi+\frac{1}{2}\hat H\wedge *\hat H\right],
\end{equation}
where $\kappa_{10}$ is the 10-dimensional Planck constant. Gauge field terms only arise at order $\alpha'$ and do, therefore, not appear in the above action. In the present paper, we restrict our discussion to the lowest order in $\alpha'$ so we will not consider gauge fields explicitly from hereon.

For completeness, we also provide the bosonic equations of motion which follow from the action~\eqref{action10d}. They are given by
\begin{eqnarray}\label{Einstein}
	&\hat{R}_{MN}-\frac{1}{4}\hat H_{PQM}\hat H^{PQ}_{\phantom{PQ}N}+2\nabla_M\partial_N\hat\phi=0,\\
	&\nabla_M\left(e^{-2\hat\phi}\hat H^M_{\phantom{M}PQ}\right)=0,\\
	&\nabla^2\hat\phi-2\hat{G}^{MN}\partial_M\hat\phi\partial_N\hat\phi+\frac{1}{12}\hat H_{MNP}\hat H^{MNP}=0\; ,
\end{eqnarray}
where $\nabla$ is the covariant derivative associated to the Levi-Civita connection of $\hat{G}$. 

The fermionic partners of the bosonic fields above are the gravitino $\psi_M$, the dilatino $\lambda$ and the gauginos $\chi$, all of which are 10-dimensional Majorana-Weyl spinors. (For our spinor conventions, see Appendix~\ref{Conventions}.) Their supersymmetry transformations  are given by
\begin{eqnarray}\label{KSE10d1}
	\delta\psi_M&=&\left(\nabla_M+\frac{1}{8}{\cal\hat H}_M\right)\epsilon,\\\label{KSE10d2}
	\delta\lambda&=&\left(\not\!\nabla\hat\phi+\frac{1}{12}{\cal\hat H}\right)\epsilon,\\\label{KSE10d3}
	\delta\chi&=&F_{MN}\Gamma^{MN}\epsilon,
\end{eqnarray}
where $\epsilon$ is a 10d Majorana-Weyl spinor parametrizing the transformations. Here and in the following we use the short-hand notation 
${\cal\hat H}_M=\hat{H}_{MNP}\Gamma^{NP}$ and ${\cal\hat H}=\hat{H}_{MNP}\Gamma^{MNP}$ for the contraction of the field strength $\hat{H}$ with products of 10-dimensional gamma matrices $\Gamma^M$. For later purposes, it is useful to introduce the connection 
\begin{equation}\label{Bismut}
	\nabla^{(H)}_M\equiv\nabla_M+\frac{1}{8}{\cal\hat H}_M\; ,
\end{equation}
which appears on the right-hand side of Eq.~\eqref{KSE10d1}. This is a connection with torsion given by the NS-NS field $\hat H$. 

In this paper, we are interested in finding solutions to the Killing spinor equations $\delta\psi_M=0$, $\delta\lambda=0$ and $\delta\chi=0$. It is known that such solutions also solve the bosonic equations of motion provided that the equation of motion and the Bianchi identity for $\hat{H}$ are satisfied. As mentioned above, we will only work to lowest order in $\alpha'$ where gauge fields do not appear so we are only concerned with the Killing spinor equations $\delta\psi_M=0$ and $\delta\lambda=0$ and the Bianchi identity in its simple form~\eqref{BI}. This concludes our basic set-up and we would now like to discuss the class of solutions we will be interested in.\\

\subsection{Solution Ansatz}
It is known for a long time~\cite{Candelas:1985en}, and has been the basis of much of heterotic string phenomenology, that a direct product of four-dimensional Minkowski space with a Calabi-Yau three-fold solves the Killing spinor equations provided the dilaton $\hat\phi$ is constant and the flux $\hat H$ vanishes. Such Calabi-Yau solutions preserve four out of the 16 supercharges, corresponding to four-dimensional $N=1$ supersymmetry. 

A more general class of solutions, also preserving four supercharges, was subsequently considered by Strominger~\cite{Strominger:1986uh}. These solutions allow for a warped product between the internal six-dimensional space and four-dimensional Minkowski space as well as a non-constant dilaton and non-vanishing flux. It turns out that the warp factor in those solutions is proportional to the dilaton. Moreover, the internal manifolds are no longer restricted to be Calabi-Yau but can be more general complex, non-Kahler manifolds with $SU(3)$ structure. For later comparison it is useful to describe these manifolds using the by-now well-established classification of $SU(3)$ structures~\cite{ChiossiSalamon, LopesCardoso:2002hd} in terms of the five torsion classes $W_1,\ldots ,W_5$. (See Appendix~\ref{Torsion classes} for a brief introduction to $G$-structures and torsion classes.) In this language, Strominger's manifolds are characterized by torsion classes satisfying~\cite{LopesCardoso:2002hd, Benmachiche:2008ma}
\begin{equation}\label{Strominger}
	W_1=W_2=0\; ,\quad W_4=\frac{1}{2}W_5=d\hat{\phi}\; ,
\end{equation}
and are arbitrary otherwise. The first of these conditions implies that the manifolds are indeed complex. 

The class of heterotic solutions described by Strominger is the most general one if one insists on a maximally symmetric four-dimensional space-time and four preserved supercharges. In this paper, we will relax both of these conditions. We will ask for only two preserved supercharges and allow four-dimensional space-time to be a domain wall solution, arguably the next-simplest possibility after maximal symmetry. Why are we interested in such vacua which do not conform with the usual requirement of a four-dimensional maximally symmetric space-time? The answer is related to the structure of flux compactifications. Frequently, flux on its own is not sufficient to stabilize all the moduli and additional non-perturbative effects are needed. However, such non-perturbative effects are typically incorporated at the level of the four-dimensional effective theory and are not ``visible" when solving the 10-dimensional theory. In other words, a typical 10-dimensional solution, only reflects the perturbative structure of the model. In type IIB Calabi-Yau vacua with flux the four-dimensional potential vanishes at the minimum as a consequence of the no-scale structure~\cite{Giddings:2001yu}. At the perturbative level, this leads to the existence of Minkowski vacua with flat directions. Not all compactifications allow for perturbative vacua with such a vanishing potential. An example is provided by the heterotic compactifications on half-flat manifolds considered in Ref.~\cite{Gurrieri:2004dt, Gurrieri:2007jg, deCarlos:2005kh,Ali:2006gd,Ali:2007ra}. In this case the unstabilized moduli are no longer flat directions and the simplest solution to the four-dimensional theory at the perturbative level is expected to be a half-BPS domain wall. The main purpose of the present paper is to find the full 10-dimensional solutions which correspond to such compactifications and identify the class of internal $SU(3)$ structure manifolds which can arise in this context. We will also study in detail the relation of this 10-dimensional solution to the four-dimensional domain wall solution which arises in the compactified theory.

With this motivation in mind we now explain our Ansatz. We consider 10-dimensional metrics of the form
\begin{equation}\label{ansatz10dgeneral}
	ds_{10}^2=e^{2A(x^m)}\left(\eta_{\alpha\beta}dx^\alpha dx^\beta+e^{2\Delta(x^m)}dx^3dx^3+g_{uv}(x^m)dx^udx^v\right),
\end{equation}
where $\eta_{\alpha\beta}$ is the $2+1$-dimensional Minkowski metric, $g_{uv}$ is an internal six-dimensional metric on a compact manifold $\hat X$ and $A$ and $\Delta$ are warp factors. We have introduced three-dimensional indices $\alpha,\beta,\ldots=0,1,2$, seven-dimensional indices $m,n,\ldots=3,...9$ and six-dimensional indices $u,v,\ldots=4,...,9$ (see Appendix \ref{Conventions} for a summary of conventions). The four-dimensional part of this metric can be interpreted as a domain wall with world-volume coordinates $x^\alpha$. The coordinate $x^3$, transverse to the domain wall, will also be denoted by $y$ in the following. The full 10-dimensional metric represents a warped product between this domain wall and the ``internal" six-dimensional space $\hat{X}$. Alternatively, the metric can be viewed as a warped product of $2+1$-dimensional Minkowski space (the world-volume of the domain wall) and a seven-dimensional space $Y=\{y\}\times \hat{X}$. Both viewpoints will be useful.

We would like to preserve $2+1$ dimensional Lorentz invariance on the domain wall world volume and therefore demand that~\footnote{This requirement still allows a ``space-filling" three form $H_{\alpha\beta\gamma}$ on the domain wall. In view of the envisaged connection with flux compactifications to four dimensions we will not consider this possibility. For the same reason we will later set $\hat{H}_{3MN}=0$.} 
\begin{equation}
 \hat{H}_{\alpha MN}=0\; ,\quad \partial_\alpha\hat{\phi}=0\; . \label{fluxansatz}
\end{equation} 
This completely specifies the Ansatz for the bosonic fields.

In addition, we should also provide the Ansatz for the spinor $\epsilon$ which parameterizes the 10-dimensional supersymmetry transformations. Since we are interested in solutions with two preserved supercharges we should assume the existence of a globally defined seven-dimensional Majorana spinor $\eta$ on $Y$. In analogy with the decomposition of the metric~\eqref{ansatz10dgeneral}, we write
\begin{equation}
 \epsilon(x^m)=\rho\otimes \eta(x^m)\otimes\theta\; ,\label{spinoransatz}
\end{equation}
where $\theta$ is an eigenvector of the third Pauli matrix $\sigma^{\underline{3}}$, and $\rho$ is a (constant) Majorana spinor in 2+1 dimensions whose two components represent the two preserved supercharges of the solution.  
In what follows it will sometimes be useful to write $\eta$ in terms of two chiral six-dimensional spinors $\eta_\pm$ as
\begin{equation}
 \eta(x^m)=\frac{1}{\sqrt{2}}\left(\eta_+(x^m)+\eta_-(x^m)\right)\; . \label{etadecomp}
\end{equation}
(See Appendix~\ref{Conventions} for details on spinor conventions.)\\

Before embarking on a detailed analysis of the above Ansatz we would like to draw two simple conclusions. From the gravitino Killing spinor equation, $\delta\psi_m=0$, together with Eqs.~\eqref{KSE10d1}, \eqref{Bismut} and \eqref{spinoransatz} we have
\begin{equation}
 \nabla_m^{(H)}\eta = 0\; . \label{intKSE}
\end{equation} 
Hence, the internal spinor $\eta$ is covariantly constant with respect to the torsion connection $\nabla^{(H)}$. Further, after a short calculation, the external part of the gravitino Killing spinor equation, together with Eq.~\eqref{ansatz10dgeneral} leads to
\begin{equation}
 \delta\psi_\alpha=\frac{1}{2}{\Gamma_\alpha}^m\partial_mA\epsilon = 0\; ,
\end{equation} 
where we have used the relation $\hat{\nabla}_M=\tilde{\nabla}+\frac{1}{2}{\Gamma_M}^N\partial_NA$ between the Levi-Civita connections associated to two metrics $\hat{G}$ and $\tilde{G}$ related by a conformal re-scaling $\hat{G}=e^{2A}\tilde{G}$.
The warp factor $A$ is, therefore, constant. For convenience, we set it to zero which simplifies our metric Ansatz~\eqref{ansatz10dgeneral} to
\begin{equation}\label{ansatz10d}
	ds_{10}^2=\eta_{\alpha\beta}dx^\alpha dx^\beta+e^{2\Delta(x^m)}dy^2+g_{uv}(x^m)dx^udx^v\; . 
\end{equation}
This concludes our set-up. We will now analyze the resulting solutions using the formalism of $SU(3)$ (and $G_2$) structures, beginning with the simple case of vanishing flux and constant dilaton and, subsequently, considering the most general case.

\section{Vanishing flux and half-flat compactifications}\label{G2 holonomy}\setall

In this section, we would like to focus on the specific case of vanishing flux and constant dilaton, that is
\begin{equation}
 \hat{H}=0\; ,\quad \hat{\phi}=\mbox{constant}\; .
\end{equation}
As a first step, we will look at the structure of the 10-dimensional solution. We find that the six-dimensional internal space $\hat{X}$ is restricted to be half-flat while the structure of the four-dimensional domain wall is described by Hitchin's flow equations. These results are then related to the four-dimensional $N=1$ supergravity obtained from compactification on half-flat mirror manifolds.
In particular, within these four-dimensional effective supergavity theories, we find an explicit half-BPS domain wall solution which precisely matches the domain wall present in the 10-dimensional solution. This shows that heterotic compactifications on half-flat mirror manifolds are indeed consistent in the sense of there being an associated solution of the full 10-dimensional theory, something taken on faith in earlier papers~\cite{Gurrieri:2004dt, Gurrieri:2007jg, deCarlos:2005kh}. 

\subsection{The 10-dimensional solution} \label{d10sol}
In the absence of flux the internal gravitino Killing spinor equation reads
\begin{equation}
 \nabla_m\eta =0 \; ,
\end{equation}
where we recall that $\nabla$ is the ordinary Levi-Civita connection. Hence, $\eta$ is a covariantly constant spinor on the seven-dimensional space $Y$. By a well-known argument this implies that $Y$ has holonomy $G_2$ (or smaller) and that its metric must be Ricci-flat. Of course, it is immediately clear that, in the absence of stress energy, a product of 2+1-dimensional Minkowski space and a seven-dimensional manifold with $G_2$ holonomy solves the 10-dimension Einstein equation~\eqref{Einstein}. 

We can also describe this situation in terms of $G_2$ structures on $Y$. (For a brief review on $G$ structures and torsion classes see Appendix~\ref{Torsion classes}.) We can think of such a $G_2$ structure as being defined by a three-form $\varphi=\frac{1}{6}\varphi_{mnp}dx^m\wedge dx^n\wedge dx^p$ and its (seven-dimensional) Hodge dual $\Phi=*_{7}\varphi$ on $Y$. In terms of the spinor $\eta$ these forms can be written as
\begin{equation}\label{3form}
	\varphi_{mnp}=-i \eta^\dagger\gamma_{mnp}\eta\; ,\quad
	\Phi_{mnpq}=\eta^\dagger\gamma_{mnpq}\eta.
\end{equation}
Then, the space $Y$ has holonomy $G_2$ (or smaller) if and only if the $G_2$ structure is torsion-free, that is, if it satisfies
\begin{equation}
 d_7\varphi=d_7\Phi=0\; , \label{notorsion}
\end{equation} 
where $d_7$ is the seven-dimensional exterior derivative. We would now like to decompose these equations into $6+1$ dimensions in accordance with our metric Ansatz~\eqref{ansatz10d}. First, in the direction of the special coordinate $y$, we  introduce the one-form
\begin{equation}
 \alpha = e^\Delta dy
\end{equation} 
satisfying
\begin{equation}
 d\alpha=\Theta\wedge\alpha\; , \quad\Theta=d\Delta\; .
\end{equation} 
In terms of the six-dimensions chiral spinors $\eta_\pm$ one can introduce the forms
\begin{equation}
 J_{uv}=\mp i\eta_\pm^\dagger\gamma_{uv}\eta_\pm\;, \quad \Omega_{uvw}=\eta_+^\dagger\gamma_{uvw}\eta_-\; ,
\end{equation} 
which define an $SU(3)$ structure on the six-dimensional space $\hat{X}$ for every fixed value of $y$. 
The definition of the $G_2$ structure~\eqref{3form} and the spinor decomposition~\eqref{etadecomp} then lead immediately to the well-known relations
\begin{equation}\label{phireduction}
	\varphi=\alpha\wedge J+\Omega_-\; ,\quad\Phi=\alpha\wedge\Omega_++\frac{1}{2}J\wedge J\; ,
\end{equation}
where $\Omega_\pm$ are the real and imaginary parts of $\Omega$. These relations express the $G_2$ structure on $Y$ in terms of the $SU(3)$ structure on the six-dimensional space $\hat{X}$ and the one-form $\alpha$ in the $y$ direction. They can be used to rewrite the vanishing torsion conditions~\eqref{notorsion} for the $G_2$ structure as
\begin{eqnarray}\label{HF1}
	d\Omega_-&=&0,\\\label{HF2}
	J\wedge\ dJ&=&0,\\\label{Hitchin1Theta}
	d\Omega_+&=&e^{-\Delta} J\wedge\partial_{y} J-\Theta\wedge \Omega_+,\\\label{Hitchin2Theta}
	dJ&=&e^{-\Delta}\partial_{y}\Omega_--\Theta\wedge J.
\end{eqnarray}
The first two of these equations imply that the $SU(3)$ structure on the six-dimensional space $\hat X$ is, in fact, half-flat. A half-flat $SU(3)$ structure can also be characterized by the following conditions
\begin{equation}\label{Jtorsion}
W_{1-}=W_{2-}=W_4=W_5=0\; ,
\end{equation}
on the torsion classes, as can be seen by comparison with the general expressions for $dJ$ and $d\Omega$ in terms of torsion classes, Eqs.~\eqref{su3torsion}. Here and in the following, we use subscripts $\pm$ to denote the real and imaginary parts of torsion classes. Note that, unlike for Strominger's class of solutions~\eqref{Strominger}, $W_1$ and $W_2$ are non-zero in general and, hence, the manifold $\hat{X}$ does not necessarily admit an integrable complex structure. A further comparison between Eqs.~\eqref{Hitchin2Theta} and \eqref{Jtorsion} reveals that
\begin{equation}
 \Theta=0\; .
\end{equation}
Hence, the warp factor $\Delta$ can be set to zero and the 10-dimensional string-frame metric~\eqref{ansatz10d} takes the form
\begin{equation}
 ds^2_{10}=\eta_{\alpha\beta}dx^\alpha dx^\beta+dy^2+g_{uv}(y,x^w)dx^udx^v\; , \label{hfmetric}
\end{equation}
where $g_{uv}$ is the metric associated to the half-flat $SU(3)$ structure given by $J$ and $\Omega$. From Eqs.~\eqref{Hitchin1Theta} and \eqref{Hitchin2Theta} the $y$ dependence of this $SU(3)$ structure is described by Hitchin's flow equations~\cite{ChiossiSalamon}
\begin{equation}
 d\Omega_+=J\wedge\partial_yJ\; ,\quad dJ=\partial_y\Omega_-\; . \label{hfe}
\end{equation} 
From a physics point of view, the metric~\eqref{hfmetric} should be interpreted as a product of a six-dimensional half-flat space $\hat{X}$ with metric $g_{uv}$ and a four-dimensional domain wall with world-volume coordinates $x^\alpha$ and transverse direction $y$. This shows that half-flat spaces can indeed be considered as solutions of the heterotic string provided they are ``paired up" with an external domain wall solution rather than a maximally symmetric four-dimensional space-time. The existence of these solutions also justifies heterotic compactifications on half-flat manifolds, as carried out in Refs.~\cite{Gurrieri:2004dt, Gurrieri:2007jg, deCarlos:2005kh}, and suggests the existence of half-BPS domain wall solutions in the associated four-dimensional $N=1$ supergravity theories which should match the domain wall part of the metric~\eqref{hfmetric}. We will now verify this picture explicitly for compactifications on half-flat mirror manifolds. First, we will briefly review the four-dimensional $N=1$ supergravity theories which originate from such compactifications. Then we find explicit half-BPS domain wall solutions in these supergravity theories and show that they match the 10-dimensional solutions just obtained.

\subsection{Heterotic compactification on half-flat mirror manifolds}\label{SUGRA}

Half-flat mirror manifolds have first been introduced in the context of type II mirror symmetry with NS-NS flux~\cite{Gurrieri:2002wz}. Essentially, they arise as mirrors of type II Calabi-Yau compactifications with electric NS-NS flux. More specifically, consider a mirror pair $X$, $\tilde{X}$ of Calabi-Yau manifolds and compactification of type IIB string theory on $\tilde{X}$ with NS-NS flux $\tilde{H}=e_i\tilde{\beta}^i$, where $i=1,\ldots, h^{2,1}(\tilde{X})$, the $\tilde{\beta}^i$ are part of the standard symplectic three-form basis on $\tilde{X}$ and $e_i$ are integer flux parameters. Then mirror symmetry suggests the existence of a half-flat manifold $\hat{X}$, closely related to the mirror Calabi-Yau manifolds $X$, so that compactification of IIA on $\hat{X}$ (without flux) is mirror to the IIB compactification on $\tilde{X}$ with flux $\tilde{H}$. Manifolds $\hat{X}$ of this type will be referred to as half-flat mirror manifolds. 

Mirror symmetry allows one to describe a number of properties of a half-flat mirror manifold $\hat{X}$ which, in turn, facilitates explicit compactifications on such manifolds. Usually, these properties can be formulated in terms of related properties of the associated Calabi-Yau manifold $X$. In particular, $\hat{X}$ carries a set $\{\omega_i\}$, where $i,j,\ldots = 1,\ldots , h^{1,1}(X)$ of two forms and a symplectic basis $\{\alpha_A,\beta^A\}$, where $A,B,\ldots = 0,\ldots , h^{2,1}(X)$, of three-forms so that the $SU(3)$ structure forms $(J,\Omega)$ can be expanded as
\begin{equation}\label{Jexpansion}
	J=v^i\omega_i\; ,\quad\Omega={\cal Z}^A\alpha_A-\mathcal{G}_A\beta^A\; .
\end{equation}
These equations are in complete analogy with the expansion of the Kahler form and holomorphic three-form on a Calabi-Yau manifold and, hence, by abuse of terminology, we will also refer to the $v^i$ and ${\cal Z}^A$ as Kahler and complex structure moduli, respectively. We also introduce the affine complex structure moduli $z^a={\cal Z}^a/{\cal Z}^0$, where $a,b,\ldots =1,\ldots ,h^{2,1}(X)$. Many of the standard Calabi-Yau moduli space results apply and the ones relevant in the present context are summarized in Appendix~\ref{Moduli space geometry}. For a non-Calabi-Yau manifold $J$ and $\Omega$ are no longer closed and given the above expansion the same must be true for at least some of the forms $\{\omega_i\}$ and $\{\alpha_A,\beta^A\}$. It turns out, in the present case, the only non-closed forms are~\cite{Gurrieri:2002wz}
\begin{equation}
d\omega_i=e_i\beta^0\; ,\quad d\alpha_0=e_i\tilde{\omega}^i\; . \label{hfmdef}
\end{equation}
Here $\{\tilde{\omega}^i\}$ is a set of four-forms dual to $\{\omega_i\}$, so that
\begin{equation}
  \int \omega_i\wedge\tilde\omega^j=\delta_i^j\; .
\end{equation}
With these relations it is easy to verify that
\begin{equation}
	dJ=v^ie_i\beta^0\; ,\quad d\Omega={\cal Z}^0e_i\tilde\omega^i\; , \label{dJ}
\end{equation}
and that $J$ and $\Omega$ indeed satisfy the half-flat conditions~\eqref{HF1} and \eqref{HF2}.\\

Heterotic compactifications on half-flat mirror manifolds have been studied in Ref.~\cite{Gurrieri:2004dt} and here we briefly review the main results. We begin with the reduction Ansatz and the relation between the 10- and four-dimensional fields. The six-dimensional internal space is taken to be the half-flat mirror space $\hat{X}$ with the metric $g_{uv}$ associated to the $SU(3)$ structure $(J,\Omega)$. In terms of the total internal volume ${\mathcal{V}=\int d^6x\sqrt{g}}$ the four-dimensional dilaton $\phi$ is given by
\begin{equation}\label{dilaton}
	\phi=\hat{\phi}-\frac{1}{2}{\rm ln}\mathcal{V},
\end{equation}
where $\hat{\phi}=\hat{\phi}(x^\mu)$ is the zero mode of the 10-dimensional dilaton. The Ansatz for the 10-dimensional metric then reads
\begin{equation}\label{ansatzHF}
	ds_{10}^2=e^{2\phi}g_{4\mu\nu}dx^\mu dx^\nu+g_{uv}dx^udx^v,
\end{equation}
where the dilaton factor in front of the four-dimensional part has been included so that $g_{4\mu\nu}$ is the four-dimensional Einstein-frame metric. Further, the zero-mode expansion of the NS-NS field is
\begin{equation}
 \hat{B}=B+b^i\omega_i\; , \quad\hat{H}=H+db^i\wedge\omega_i+b^id\omega_i\label{Bzero}
\end{equation}
where $b^i$ are axionic scalars and $B=\frac{1}{2}B_{\mu\nu}dx^\mu\wedge dx^\nu$ is a four-dimensional two-form with field strength $H=dB$ which can be dualized to the universal axion $a$. Note that, even thought we are considering the case without ``explicit" flux, a non-zero flux is induced from the last term in Eq.~\eqref{Bzero} as a consequence of the differential relations~\eqref{hfmdef} for half-flat mirror manifolds.
These various scalar fields form the lowest components of four-dimensional chiral supermultiplets in the usual way, that is
\begin{equation}
 S=a+ie^{-2\phi}\; ,\quad T^i=b^i+iv^i\;, \quad Z^a=z^a\; . \label{superfields}
\end{equation} 
Their Kahler potential is given by the same expression as for Calabi-Yau compactifications, namely
\begin{equation}
 K=-\ln(i(\bar{S}-S))+K^{(1)}+K^{(2)}\; ,\quad K^{(1)}=-\ln(8{\cal V})\; ,\quad K^{(2)}=-\ln\left(i\int_{\hat{X}}\Omega\wedge\bar{\Omega}\right)\; .
 \label{hfk}
\end{equation} 
Some standard results on the explicit moduli dependence of $K$ and related issues are summarized in Appendix~\ref{Moduli space geometry}. The superpotential can be obtained from the Gukov-Vafa type formula~\cite{GVW}
\begin{equation}
 W=\sqrt{8}\int_{\hat{X}}\Omega\wedge(\hat{H}+idJ)\; . \label{hfw}
\end{equation} 
For half-flat mirror manifolds and vanishing flux this superpotential has the explicit form
\begin{equation}
 W=\sqrt{8}e_iT^i\; ,
\end{equation}
where we have used the relations~\eqref{dJ} and \eqref{Bzero}. Even though we are not considering explicit flux the $\hat{H}$ term has to be included in this formula to correctly incorporate the flux induced by the structure of the half-flat mirror manifolds (see Eq.~\eqref{Bzero}).

\subsection{Four-dimensional domain wall solutions}\label{4d geometry}

We would now like to find explicit half-BPS domain solutions within the four-dimensional $N=1$ supergravity theories just discussed. As a preparation, we first set up the general formalism for four-dimensional BPS domain walls (see Refs.~\cite{Cvetic:1996vr, Wess:1992cp, Eto:2003bn} for further details).

Consider a four-dimensional N=1 supergravity theory with chiral superfields $(A^I,\chi^I)$, Kahler potential $K$ and superpotential $W$ and a gravitino $\psi_\mu$. Then, the Killing spinor equations are given by
\begin{eqnarray}\label{KSE4d1}
	\delta\chi^I&=&i \sqrt{2} \sigma^\mu \bar{\zeta} \partial_\mu A^I-\sqrt{2}e^{K/2}K^{IJ^*}D_{J^*}W^*\zeta=0,\\\label{KSE4d2}
	\delta\psi_\mu&=&2 \mathcal{D}_\mu\zeta+ie^{K/2}W\sigma_\mu\bar{\zeta}=0\; ,
\end{eqnarray}
where the Weyl spinor $\zeta$ parameterizes supersymmetry, $D_IW=W_I+K_IW$ and $(\sigma^{\underline{\mu}})=({\bf 1}_2,\sigma^{\underline{\alpha}})$, with the Pauli matrices $\sigma^{\underline{a}}$ . The covariant derivatives $\mathcal{D}_\mu$ is defined by
\begin{equation}
 \mathcal{D}_\mu=\partial_\mu+\omega_\mu+\frac{1}{4}(K_j\partial_\mu A^j-K_{j^*}\partial_\mu A^{j^*})\; ,
\end{equation} 
with the spin connection $\omega_\mu$.\\

For a domain wall solution, we should split the coordinates as $(x^\mu)=(x^\alpha ,y)$ where $\alpha,\beta,\ldots = 0,1,2$ label the directions longitudinal to the domain wall and $y$ is the transverse coordinate. Accordingly, we should start with an Ansatz
\begin{equation}
 ds_4^2=e^{-2B}\left(\eta_{\alpha\beta}dx^\alpha dx^\beta +dy^2\right) \label{4dmetric}
\end{equation}
for the metric, where $B=B(y)$ is a warp factor. In addition, all scalar fields and the spinor $\zeta$ are functions of $y$ only. For the spin connection of the metric Ansatz~\eqref{4dmetric} one has
\begin{equation}
	\omega_0=-\frac{1}{2}B '\sigma^{\underline{2}}\; ,\quad \omega_1=-i\frac{1}{2}B ' \sigma^{\underline{3}}\; ,\quad
	\omega_2=i\frac{1}{2}B '\sigma^{\underline{1}}\; ,\quad\omega_3=0\; .
\end{equation}
Here and in the following we use a prime to denote the derivative with respect to $y$. The general Killing spinor equations~\eqref{KSE4d1} and \eqref{KSE4d2} then specialize to
\begin{equation}\label{DW}
 \begin{array}{rcl}
	\partial_yA^I&=&-ie^{-B}e^{K/2}K^{IJ^*}D_{J^*}W^*,\\
	B '&=&ie^{-B}e^{K/2}W,\\
	{\rm Im} (K_I\partial_yA^I)&=&0,\\
	2\zeta '&=&-B'\zeta\; .
 \end{array}
\end{equation}
Further, the spinor $\zeta$ satisfies the constraint
\begin{equation}\label{1/2BPS}
	\zeta(y)=\sigma^{\underline{2}}\bar{\zeta}(y)\; ,
\end{equation}
which reduces the number of independent spinor components to two, corresponding to half-BPS solutions.\\

We would now like to solve the system of Killing spinor equations \eqref{DW} for the specific supergravity theories obtained from compactification on half-flat mirror manifolds, as discussed in the previous sub-section. We recall that the (chiral) field content of these theories consists of $(A^I)=(S,T^i,Z^a)$ and the Kahler potential and superpotential are given by Eqs.~\eqref{hfk} and \eqref{hfw}, respectively. In the following, we will make frequent use of the properties of the Kahler potential summarized in Appendix~\ref{Moduli space geometry}. The fields are split up into their real and imaginary part according to
\begin{equation}
 S=a+ie^{-2\phi}\; ,\quad T^i=b^i+iv^i\; ,\quad Z^a=c^a+iw^a\; . \label{STZreal}
\end{equation} 

It is not difficult to see by inspection of~\eqref{DW} that the right-hand side of the first equation is purely imaginary. This implies that the real parts of all superfields must be constant, that is
\begin{equation}
 a\sim b^i\sim c^a\sim {\rm const}\; , \quad b^ie_i=0\; , \label{imsol}
\end{equation} 
where the last constraint follows because the superpotential needs to be purely imaginary as a consequence of the second equation~\eqref{DW}. By comparing the first equations~\eqref{DW} for the dilaton $S$ with the second equation~\eqref{DW} one finds that $\phi '=B'$ so without loss of generality we can set
\begin{equation}
 B=\phi\; .
\end{equation} 
For the remaining imaginary parts we have
\begin{equation} \label{dweq}
 \phi '=-\frac{1}{2}\frac{\calk '}{\calk}\; ,\quad
 \calk_{ij}\partial_yv^j=\sqrt{\frac{\calk}{\tilde\calk}}e_i\; ,\quad
 \partial_yw^a=-2\phi' w^a\; .
\end{equation} 
Here, $\calk=\calk_{ijk}v^iv^jv^k$ is the Kahler moduli pre-potential where $\calk_{ijk}$ are the intersection numbers of $X$ and $\calk_{ij}=\calk_{ijk}v^k$. Analogously, for the complex structure module, the pre-potential is given by $\tilde{\calk}=\tilde{\calk}_{abc}w^aw^bw^c$ with the intersection numbers $\tilde{\calk}_{abc}$ of the mirror Calabi-Yau $\tilde{X}$. (For details see Appendix~\ref{Moduli space geometry}.)

These equations can easily be integrated to
\begin{equation}
 \calk=\calk_0e^{-2\phi}\; ,\quad \calk_i=2\sqrt{\frac{\calk_0}{\tilde{\calk}({\bf k})}}e_i\tilde{y}+\calk_{0i}\; ,\quad w^a=k^a e^{-2\phi}\; , \label{sol}
\end{equation} 
where we have introduced a new coordinate $\tilde{y}$ defined by $d\tilde{y}=e^{2\phi}dy$. Further, $\calk_0$, $\calk_{0i}$ and $k^a$ are integration constants and $\tilde{\calk}({\bf k})$ denotes the complex structure pre-potential as a function of the $k^a$. To find the explicit solution in terms of the Kahler moduli $v^i$ one has to invert the relations
\begin{equation}
 \calk_i=\calk_{ijk}v^jv^k\; ,
\end{equation}
which can only be done on a case by case basis. 

\subsection{Comparison between 10 and four dimensions} \label{comp}

We would now like to show that this four-dimensional domain wall indeed matches our 10-dimensional solution~\eqref{hfmetric}, in a way similar to what happens in the context of type IIA~\cite{Mayer:2004sd, Smyth:2009fu}. We start by re-writing the four-dimensional domain wall Killing spinor equations~\eqref{dweq} in terms of 10-dimensional language. To this end, we insert the following definitions of the 10-dimensional fields,
\begin{equation}
	\calk=({\cal Z}^0)^2\tilde\calk\; ,\quad \hat{\phi}'=\phi'+\frac{1\calk'}{2\calk}.
\end{equation}
The first of these arises from the compatibility relation~\eqref{su3def} while the latter is simply the definition~\eqref{dilaton}. It is then straightforward to see that the four-dimensional domain wall equations~\eqref{dweq} are equivalent to the following,
\begin{equation}\label{DWmatch}
	\hat\phi'=0\;,\quad \calk_{ij}\partial_yv^j={\cal Z}^0 e_i\; ,\quad {\cal Z}^0\omega^a={\rm const}\;.
\end{equation}
It is also useful to recall from Eq.~\eqref{imsol} the constraints these equations imply on the real parts of the superfields, namely
\begin{equation}\label{Realpartmatch}
	a\sim b^i\sim c^a\sim {\rm const}\; , \quad b^ie_i=0\; .
\end{equation}
\\

We now turn to the 10-dimensional solution~\eqref{hfmetric} and will show that it corresponds to the above system. To do this, we insert the defining relations of mirror half-flat manifold~\eqref{dJ} into Hitchin's flow equations~\eqref{hfe}. We can easily see that the first flow equation gives,
\begin{equation}\label{viflow}
	\calk_{ij}\partial_yv^j={\cal Z}^0 e_i,
\end{equation}
which is equivalent to the corresponding domain wall equation~\eqref{DWmatch}. From the second Hitchin flow equation, we obtain three equations,
\begin{equation}
	{\cal Z}^0\omega^a={\rm const}\;,\quad c^a={\rm const}\;,\quad v^ie_i=\frac{1}{6}({\cal Z}^0\tilde\calk)' \;,
\end{equation}
which correspond to the components of the three-forms $\alpha_a$, $\beta^a$ and $\beta^0$, respectively. The first two equations are identical equations in~\eqref{DWmatch} and~\eqref{Realpartmatch}. The third one simply tells us that the flow equations preserves the compatibility relation~\eqref{su3def} between the SU(3) structure forms $J$ and $\Omega$. This last equation does not provide any new information as it is just a contracted version of~\eqref{viflow} together with the condition ${\cal Z}^0\omega^a={\rm const}$.

Finally, we need to realize that the last conditions in~\eqref{DWmatch} and~\eqref{Realpartmatch} ensure, from a 10-dimensional point of view, the vanishing of $\hat H$ and a constant dilaton $\hat \phi$. Therefore, we have shown that the four-dimensional domain wall solution of the effective supergravity is completely equivalent to the corresponding 10-dimensional flow equations.

\section{Non-vanishing flux and half-flat compactifications}\label{Adding flux}\setall

We will now extend the discussion of the previous section by including non-vanishing NS-NS flux, as well as a non-constant dilaton. First, we derive the generalization of Hitchin's flow equations for this case and then discuss the relation to domain wall solutions in the four-dimensional effective supergravity compactified on mirror half-flat manifolds.

\subsection{The 10-dimensional solution} \label{10dsolution}
As before, we begin by working out the constraints on the $G_2$ structure of the seven-dimensional space $Y$. 
Starting point is the seven-dimensional part of the gravitino Killing spinor equation and the dilatino Killing spinor equation. From \eqref{KSE10d1} and \eqref{KSE10d2} they read
\begin{equation}
 \nabla_m\eta=-\frac{1}{8}{\cal H}_m\eta\; ,\quad \not\!\nabla\hat\phi\,\eta=-\frac{1}{12}{\cal\hat H}\,\eta\; .
\end{equation} 
We proceed in the usual way by multiplying the above equations and their hermitian conjugates with anti-symmetrised products of gamma matrices times $\eta$ or $\eta^\dagger$. With the definitions~\eqref{3form} of the $G_2$ structure forms $\varphi$ and $\Phi$, this leads to the following set of equations
\begin{eqnarray}
 \nabla_{m}\varphi_{npq}&=&\frac{3}{2}\hat H_{ms[n}{\varphi_{pq]}}^s\nn\\
 \nabla_{m}\Phi_{npqr}&=&-2\hat H_{ms[n}{\Phi_{pqr]}}^s\nn\\
 \nabla_{[m}\hat\phi\,\Phi_{npqr]}&=&\hat H_{s[mn}{\Phi_{pqr]}}^s\\
 4\nabla_{[m}\hat\phi\,\varphi_{npq]}&=&-3\hat H_{v[mn}{\varphi_{pq]}}^v+\frac{1}{12}\epsilon_{mnpqrst}\hat H^{rst}\nn\\
 \epsilon_{mnpqrst}\nabla^t\hat\phi&=&-10\hat H_{[mnp}\varphi_{qrs]}\nn\\
 \hat H_{[mnp}\Phi_{qrst]}&=&0\; .\nn
\end{eqnarray} 
A combination of the first with the third equation, the second with the fourth equation and the last equations on their own can then be written in the form~\cite{Friedrich, Gauntlett:2002sc, Gran:2005wf}
\begin{eqnarray}
 d_7\varphi&=&2 d_7\hat\phi\wedge \varphi-*_7\,\hat H\nn\\
 d_7\Phi&=&2d_7\hat\phi\wedge \Phi\label{g2gen}\\
 *_7\,d_7\hat\phi&=&-\frac{1}{2}\hat H\wedge\varphi\nn\\
 0&=&\hat H \wedge\Phi\; ,\nn
\end{eqnarray} 
where $*_7$ and $d_7$ are the seven-dimenionsal Hodge star and exterior derivative, respectively. The first two of these equations characterize the $G_2$ structure on $Y$ and are the generalization of the torsion-free conditions~\eqref{notorsion} which appeared in the case without flux. The third equation determines the variation of the dilaton. 

From these results we can deduce the structure of $G_2$ torsion classes ${\cal X}_1,\ldots ,{\cal X}_4$. By comparing with the general results \eqref{g2torsion}, it follows that ${\cal X}_1={\cal X}_2=0$, the class ${\cal X}_3$ is determined by the corresponding component of the flux $\hat{H}$ and the class ${\cal X}_4$ by the derivative $d_7\hat{\phi}$ of the dilaton and the corresponding component of $\hat{H}$. This means that the $G_2$ structure is integrable conformally balanced~\cite{Friedrich}.\\

As previously, we now split these equations up into $6+1$ dimensions and express them in terms of the $SU(3)$ structure on $\hat{X}$. Since we are motivated by compactifications to four-dimensions and to simplify matters we set all remaining components of the flux breaking four-dimensional Lorentz symmetry to zero, that is,
\begin{equation}\label{Hy0}
 \hat{H}_{ymn}=0\; .
\end{equation}
We recall that the relation between $G_2$ and $SU(3)$ is given by Eq.~\eqref{phireduction}. Using these relations in \eqref{g2gen} and splitting up the resulting equations into a one- and a six-dimensional part we find
\begin{eqnarray}\label{dO-H}
 d\Omega_-&=&2d\hat\phi\wedge\Omega_-\\\label{dJ-H}
 dJ&=&e^{-\Delta}\Omega_-'-2e^{-\Delta}{\hat\phi}'\,\Omega_-+2d\hat\phi\wedge J -J\wedge\Theta-*\hat H\\\label{JdJ-H}
J\wedge dJ&=&J\wedge J\wedge d\hat\phi\\
d\Omega_+&=&e^{-\Delta}J\wedge J'-e^{-\Delta}{\hat\phi}' J\wedge J+2d\hat\phi\wedge\Omega_+ +\Omega_+\wedge\Theta\label{dO}\\\label{HJ}
*d\hat\phi&=&-\frac{1}{2}\hat H\wedge J\\\label{beforelast}
e^{-\Delta}{\hat\phi}'\,*1&=&-\frac{1}{2}\hat H\wedge\Omega_-\\\label{last}
0&=&\hat H\wedge \Omega_+\; ,
\end{eqnarray}
where the Hodge star and the exterior derivative now refer to six dimensions. We also recall that $\Theta =d\Delta$ is the exterior derivative of the warp factor $\Delta$. Matching up the first four of these equations with the general expressions for $dJ$ and $d\Omega$ in Eq.~\eqref{su3torsion} one finds the torsion classes are constrained by
\begin{equation}
 W_{1-}=W_{2-}=0\;,\quad W_4=\frac{1}{2}W_5=d\hat{\phi}\; , \label{tc}
\end{equation} 
and arbitrary otherwise. We can compare this result with the torsion constraints~\eqref{Strominger} which characterize Strominger's class of solutions. The only difference is that $W_{1+}$ and $W_{2+}$ can now be non-zero and, as a consequence, the six-dimensional space $\hat{X}$, while still having an almost complex structure, does no longer need to be complex. Further, since $W_4$ and $W_5$ are non-vanishing and proportional to the dilaton the $SU(3)$ structure is mildly more general than that for half-flat manifolds. We refer to this structure as generalized half-flat.

Given that we have already fixed $W_5$ in terms of the dilaton, a comparison between Eq.~\eqref{dO} and Eq.~\eqref{su3torsion} reveals that $\Theta =0$. Hence, without loss of generality we can set the warp factor $\Delta$ to zero and, as before, the 10-dimensional metric for our solution is
\begin{equation}
 ds^2_{10}=\eta_{\alpha\beta}dx^\alpha dx^\beta+dy^2+g_{uv}(y,x^w)dx^udx^v\; . 
\end{equation}
Here, $g_{uv}$, for every value of $y$, is the metric associated to the $SU(3)$ structure with torsion classes satisfying~\eqref{tc} and with $y$-dependence governed by
\begin{eqnarray}\label{10dHflow1}
	d\Omega_+=J\wedge J'-{\hat\phi}' J\wedge J+2d\hat\phi\wedge\Omega_+ \\\label{10dHflow2}
	dJ=\Omega_-'-2{\hat\phi}'\Omega_-+2d\hat\phi\wedge J-*\hat H\; .
\end{eqnarray}
These are the generalizations of Hitchin's flow equations~\eqref{hfe} in the presence of non-zero NS-NS flux. The flux in this solution must be a harmonic form on the six-dimensional space $\hat{X}$, that is,
\begin{equation}
 d\hat H=0\; ,\quad d*\hat H=0\; .
\end{equation} 

As a consistency check we can assume that all fields are $y$-independent. In this case we indeed recover the standard equations for solutions with four supercharges and maximally symmetric four-dimensional space~\cite{Held:2010az}, as we should.

\subsection{Four-dimensional domain wall solution}\label{4dDWH}

We would now like to discuss the above solutions from the viewpoint of the four-dimensional supergravity theory. In section~\eqref{SUGRA} we have reviewed the structure of this four-dimensional theory for the case of compactifications on half-flat manifolds $\hat{X}$ with vanishing $\hat H$ flux. Here, we want to show that the results still apply when this assumption is relaxed. \\

We will then continue to assume that the internal manifold is described by the mirror half-flat properties.
Hence, the superfields of the four-dimensional supergravity theory are still given by $\{S,T^i,Z^a\}$ and they are related to their higher-dimensional counterparts as in Eq.~\eqref{superfields}. The NS-NS zero-mode expansion is now given by,
\begin{equation}\label{Hfluxexpansion}
	\hat H=H+db^i\wedge\omega_i+b^id\omega_i+\epsilon_a\beta^a+\mu^a\alpha_a,
\end{equation}
where we introduce the electric flux $\epsilon_a$ and the magnetic flux $\mu^a$. The Kahler potential remains the standard one as given in Eq.~\eqref{hfk}. However, the superpotential is modified since it now contains the additional contribution due to flux. It can still be obtained from the heterotic Gukov formula~\eqref{hfw} which gives~\cite{Gurrieri:2004dt}
\begin{equation}
  W=\sqrt{8}\left(e_iT^i+\epsilon_aZ^a+\mu^a\mathcal{G}_a(Z)\right)\; .
\end{equation}
Here, $\mathcal{G}_a(Z)$ are the derivatives of the pre-potential (see Appendix~\ref{Moduli space geometry}) and we have set ${\cal Z}^0=1$, for simplicity. \\

To find domain wall solutions we can follow the same general set-up as in sub-section~\ref{4d geometry}, that is, we can start with the general domain wall Killing spinor equations~\eqref{DW}. As before, we split the fields into real and imaginary parts as
\begin{equation}
 S=a+ie^{-2\phi}\; ,\quad T^i=b^i+iv^i\; ,\quad Z^a=c^a+iw^a\; .
\end{equation}
In much the same way as in sub-section~\ref{4d geometry} we can conclude that the warp factor $B$ in the metric Ansatz~\eqref{4dmetric} is determined by the dilaton, so $B=\phi$ and that the real parts of the fields are subject to the constraints
\begin{equation}
 a\sim b^i\sim \mbox{const}\; ,\quad \partial_yc^a=-\sqrt{\frac{\tilde{\mathcal{K}}}{\mathcal{K}}}\mu^a\;,\quad \tilde{\mathcal{K}}_a\mu^a=0\;, \quad b^ie_i+\epsilon_ac^a=\frac{1}{2}\tilde{\mathcal{K}}_{abc}c^ac^b\mu^c\; .
\end{equation}
For the imaginary parts, we find  
\begin{eqnarray} \label{DWflux}
	-\frac{1}{4}\left(\frac{\mathcal{K}'}{\mathcal{K}}+\frac{\tilde{\mathcal{K}}'}{\tilde{\mathcal{K}}}\right)&=&2\phi'\; ,\\
	\mathcal{K}_i\left(\frac{\mathcal{K}'}{\mathcal{K}}+2\phi'\right)-\mathcal{K}_i'&=&-2\sqrt{\frac{\mathcal{K}}{\tilde{\mathcal{K}}}}e_i\; ,\\
	\tilde{\mathcal{K}}_a\left(\frac{\tilde{\mathcal{K}}'}{\tilde{\mathcal{K}}}+2\phi'\right)-\tilde{\mathcal{K}}_a'&=&-2\sqrt{\frac{\tilde{\mathcal{K}}}{\mathcal{K}}}\left(\epsilon_a-\tilde{\mathcal{K}}_{abc}c^b\mu^c\right)\; .
\end{eqnarray}
It is easy to see that, for vanishing $\epsilon_a$ and $\mu^a$ fluxes, these equations reduce to the previous ones~\eqref{dweq}.

\subsection{Comparison between 10 and four dimensions}\label{10dH}

As before, we would like to show that this four-dimensional domain wall indeed matches our 10-dimensional solution. For clarity, let us rewrite the relevant $7$-dimensional Killing spinor equations~\eqref{dO-H}-\eqref{last}. First, we should note that the relations for mirror half-flat manifolds~\eqref{hfmdef} for the basis forms $\omega_i$ and $\alpha_0$ together with Eqs.~\eqref{dO-H}, \eqref{JdJ-H} and~\eqref{HJ} imply that
\begin{equation}
	db^i=d\hat\phi=0.
\end{equation}
Let us also remind that the warp factor can be set to zero $\Delta=0$. Therefore, we are left with the Killing spinor equations
\begin{eqnarray}
	\Omega_-'&=&2\hat\phi'\Omega_-+dJ+*\hat H\;,\label{10dHeq3}\\
	J\wedge J'&=&\hat\phi'J\wedge J+d\Omega_+\;,\label{10dHeq4}\\
	\Omega_-\wedge\hat H&=&2\hat\phi'*1\;,\label{10dHeq2}\\
	\Omega_+\wedge\hat H&=&0\;.\label{10dHeq1}
\end{eqnarray}

We can now expand these equations on the basis $\{\omega_i\}$ and $\{\alpha_A,\beta^A\}$ to obtain explicit equations for the moduli fields. For this, we insert the respective expansions~\eqref{Jexpansion} and~\eqref{Hfluxexpansion} of the Kahler form $J$, the complex structure $\Omega$ and the NS-NS field $\hat H$ into the above Killing spinor equations~\eqref{10dHeq3}-\eqref{10dHeq1}. The calculations is a bite tedious due to the Hodge $*$ operator. The $\alpha_0$ component of~\eqref{10dHeq3} together with~\eqref{10dHeq1} give the two constraints,
\begin{equation}
	b^ie_i+\epsilon_ac^a=\frac{1}{2}\tilde{\mathcal{K}}_{abc}c^ac^b\mu^c\;,\quad \tilde{\mathcal{K}}_a\mu^a=0\;.
\end{equation}
The imaginary part of the complex structure and the dilaton equations are given by the $\alpha_a$ component of~\eqref{10dHeq3} together with~\eqref{10dHeq2}. We obtain,
\begin{equation}
	\frac{1}{2}\left(\frac{\partial_y{\cal Z}^0}{{\cal Z}^0}-\frac{\tilde{\mathcal{K}}'}{\tilde{\mathcal{K}}}\right)\tilde{\mathcal{K}}_a+\tilde{\mathcal{K}}_a'=\frac{2}{{\cal Z}^0}\left(\epsilon_a-\tilde{\mathcal{K}}_{abc}c^b\mu^c\right)\;,\quad 
	\hat\phi'=\frac{3\partial_y{\cal Z}^0}{4{\cal Z}^0}+\frac{\tilde{\mathcal{K}}'}{4\tilde{\mathcal{K}}}\;.
\end{equation}
The remaining components of Eq.~\eqref{10dHeq3} lead to equations for the real parts of the complex structure moduli and the last Eq.~\eqref{10dHeq4} gives the Kahler moduli $y$-dependence. These read explicitly,
\begin{equation}
	\partial_yc^a=-\frac{1}{{\cal Z}^0}\mu^a\;,\quad \frac{1}{2}\mathcal{K}_i'=\hat\phi'\mathcal{K}_i+{\cal Z}^0e_i\;.
\end{equation}
Finally, we have the conditions,
\begin{equation}
	a={\rm const}\;,\quad b^i={\rm const}\;,
\end{equation}
coming from the NS-NS flux Ansatz~\eqref{fluxansatz} and~\eqref{Hy0}.\\

Again, it is now easy to see the correspondence between the four-dimensional domain wall equations and the 10-dimensional ones. For this, we need to insert the 10-dimensional relations
\begin{equation}\label{matchrel}
	\calk=({\cal Z}^0)^2\tilde\calk\; ,\quad \hat{\phi}'=\phi'+\frac{1\calk'}{2\calk},
\end{equation}
into one or the other set of equations. It is easy to see that it will lead to equivalent equations.

\section{A class of Calabi-Yau domain wall solutions}

It is interesting to realize that our flow equations~\eqref{10dHflow1} and \eqref{10dHflow2} imply the existence of an exact solution which involves a Calabi-Yau manifold and non-vanishing $\hat H$-flux. In this solution, the flux stress-energy in the Einstein equations~\eqref{Einstein}, instead of deforming away from a Calabi-Yau space, leads to a non-trivial variation of the moduli as one moves in the direction transverse to the domain wall direction. The full seven-dimensional manifold has $G_2$ structure with a non-vanishing Ricci tensor, while, at the same time, the six-dimensional fibre at each fixed point in the coordinate $y$ remains Ricci-flat.

\subsection{10-dimensional geometry}\label{10dCY}

Calabi-Yau manifolds are characterized by the property $dJ=0$ and $d\Omega=0$. Inserting this back into the generalized flow equations~\eqref{10dHflow1} and \eqref{10dHflow2}, we find
\begin{equation}\label{CYflow}
 J\wedge J'={\hat\phi}' J\wedge J\;,\quad
 \Omega_-'=2{\hat\phi}'\,\Omega_-+*\hat H\;,\quad
 \Omega_-\wedge\hat H=2{\hat\phi}'\,*1\;.
\end{equation}
A Calabi-Yau manifold $\hat X$ with moduli varying along $y$ as dictated by the above flow equations will then be a solution of the Einstein equations~\eqref{Einstein}. To satisfy the full system of equations of motion, we also have the constraints on the flux $\hat H$ and the dilaton $\hat\phi$. From Eqs.~\eqref{dO-H}-\eqref{last} they can be written as
\begin{equation}\label{CYH}
 d\hat\phi=0\;,\quad\hat H\wedge J=0\;,\quad
 \hat H\wedge \Omega_+=0\;.
\end{equation}

We can now deploy the full range of Calabi-Yau moduli space technology to solve these differential equations for the various moduli fields. In principle, this amounts to taking the $e_i=0$ limit of our previous, general discussion in Sections~\ref{4dDWH} and~\ref{10dH}. However, for the sake of clarity, we will repeat the required steps here. We recall the standard expansion of the Kahler form and the complex structure
\begin{equation}
	J=v^i\omega_i\; ,\quad\Omega={\cal Z}^A\alpha_A-\mathcal{G}_A\beta^A\; ,
\end{equation}
as well as the expansion of the flux
\begin{equation}
	\hat H=\epsilon_a\beta^a+\mu^a\alpha_a
\end{equation}
in terms of harmonic two forms $\{\omega_i\}$ and harmonic three-forms $\{\alpha_A,\beta^B\}$ on the Calabi-Yau manifold $\hat{X}$. 
This will then satisfy the second constraint of~\eqref{CYH} from the property of the basis forms. We also recall, that we have set all components of $\hat H$ breaking four-dimensional Lorenz-invariance to zero, that is, $\hat H_{\mu MN}=0$. This implies the axions $a$ and $b^i$ have to be constant.
The $y$-dependence of the remaining moduli $v^i$ and ${\cal Z}^A$ is determined by the flow equations~\eqref{CYflow} and can be explicitly obtained by inserting the above expansions for $J$, $\Omega$ and $\hat H$ and setting the coefficient of each basis form to zero. Working in the large complex structure limit, we find for the complex structure moduli
\begin{equation}
	\partial_yc^a=-\frac{1}{{\cal Z}^0}\mu^a\;,\quad \tilde{\mathcal{K}}_a'=\frac{2}{{\cal Z}^0}\left(\epsilon_a-\tilde{\mathcal{K}}_{abc}c^b\mu^c\right)\;,
\end{equation}
and for the dilaton and the Kahler moduli
\begin{equation}
	\hat\phi'=\frac{\tilde{\mathcal{K}}'}{\tilde{\mathcal{K}}}\;,\quad \hat\phi'=\frac{\partial_y{\cal Z}^0}{{\cal Z}^0}\;,\quad \hat\phi'=\frac{\partial_yv^i}{v^i}\;.
\end{equation}
We should point out the $y$-dependence of $J$ and $\Omega$ implied by these solutions is consistent with the $SU(3)$ structure compatibility relations~\eqref{su3def}. Finally, we also have the conditions coming from the third constraint of~\eqref{CYH} together with the $\alpha_0$ component of the $\Omega_-'$ equation,
\begin{equation}\label{fluxconstraint}
	\tilde{\mathcal{K}}_a\mu^a=0\;,\quad \epsilon_ac^a=\frac{1}{2}\tilde{\mathcal{K}}_{abc}c^ac^b\mu^c\;.
\end{equation}
They are constraints on the flux parameters and the different integrations constants of the previous flow equations.\\

We can integrate these differential equations in term of the new variable $d y= {\cal Z}^0 d\tilde y$. We find for the complex structure moduli,
\begin{eqnarray}
	&c^a=-\mu^a\tilde y + \mathcal{C}^a\;,\\
	&\tilde{\mathcal{K}}_a=\frac{1}{2}\tilde{\mathcal{K}}_{abc}\mu^b\mu^c\tilde y^2+2(\epsilon_a-\tilde{\mathcal{K}}_{abc}\mu^c\mathcal{C}^b)\tilde y+\tilde{\mathcal{K}}_{0a}\;,
\end{eqnarray}
where $\mathcal{C}^a$ and $\tilde{\mathcal{K}}_{0a}$ are integration constants. This then determines the dilaton and, therefore, the Kahler moduli $v^i$ and the ${\cal Z}^0$ field:
\begin{equation}
	\tilde{\mathcal{K}}\sim e^{\hat\phi}\;,\quad   v^i\sim e^{\hat\phi}  \;,\quad {\cal Z}^0\sim e^{\hat\phi}\;.
\end{equation}
Finally, we have the constraints on the flux parameters~\eqref{fluxconstraint}. They turns out to be equivalent to the non-trivial constraint
\begin{equation}
	\tilde{\mathcal{K}}_{abc}\mu^a\mu^b\mu^c=0, \label{fluxcons}
\end{equation}
on the flux parameters $\mu^a$ and the further conditions
\begin{equation}
	(2\epsilon_a+\tilde{\mathcal{K}}_{0a})\mu^a=0\;,\quad \epsilon_a\mu^a=\frac{1}{2}\tilde{\mathcal{K}}_{abc}\mu^a\mathcal{C}^b\mathcal{C}^c\;,\quad \tilde{\mathcal{K}}_{abc}\mu^a\mu^b\mathcal{C}^c=0\;,\quad \epsilon_a\mathcal{C}^a=0\;.
\end{equation}
The first of these equation just determines the integration constants $\tilde{\mathcal{K}}_{0a}$. The other three equations can be solved by an appropriate choice of $\mathcal{C}^a$ whenever $h^{2,1}\geq 3$. Indeed, in this case, we can always find a vector $\mathcal{C}^a$ orthogonal to $\tilde{\mathcal{K}}_{abc}\mu^a\mu^b$ and $\epsilon_a$ and (possibly up to a sign issue) choose its norm so that the remaining equation is satisfied. Hence, provided that the flux satisfies the non-trivial constraint~\eqref{fluxcons} we find indeed a Calabi-Yau domain wall solution.

\subsection{Four-dimensional domain wall}

As before, we now relate this 10-dimensional Calabi-Yau domain wall solution to the four-dimensional supergravity obtained by compactifying on the Calabi-Yau manifold with flux. The module fields in this four-dimensional supergravity are
\begin{equation}
 S=a+ie^{-2\phi}\; ,\quad T^i=b^i+iv^i\; ,\quad Z^a=c^a+iw^a\; ,
\end{equation}
as usual, and the superpotential is given by
\begin{equation}
  W=\sqrt{8}\left(\epsilon_aZ^a+\mu^a\mathcal{G}_a(Z)\right)\; .
\end{equation}
In the same way as in sub-section~\ref{4d geometry}, the domain wall Killing spinor equations~\eqref{DW} tell us that the real parts of the superfields satisfy
\begin{equation}
  a=b^i=\mbox{const}\; ,\quad \partial_yc^a=-\sqrt{\frac{\tilde{\mathcal{K}}}{\mathcal{K}}}\mu^a\;,\quad   \tilde{\mathcal{K}}_a\mu^a=0 \;,\quad \epsilon_ac^a=\frac{1}{2}\tilde{\mathcal{K}}_{abc}c^ac^b\mu^c\;.
\end{equation}
Again, the warp factor of the metric Ansatz~\eqref{4dmetric} can be set to $B=\phi$. For the imaginary parts, we have
\begin{equation} 
   \partial_yv^i=-2\phi'v^i\; ,\quad
   \tilde{\mathcal{K}}_a'=2\sqrt{\frac{\tilde{\mathcal{K}}}{\mathcal{K}}}(\epsilon_a-\tilde{\mathcal{K}}_{abc}c^b\mu^c)\;,\quad
   -2\phi'=\frac{\tilde{\mathcal{K}}'}{\tilde{\mathcal{K}}}\; .
\end{equation}
We note that this corresponds to Eq.~\eqref{DWflux} in the limit where the half-flat fluxes vanish, that is, $e_i=0$. The matching of these four-dimensional flow equations with the ten-dimensional ones~\eqref{CYflow} can be worked out in the same way as previously, namely by inserting the definitions
\begin{equation}
	\calk=({\cal Z}^0)^2\tilde\calk\; ,\quad \hat{\phi}'=\phi'+\frac{1\calk'}{2\calk}.
\end{equation}
Hence, we conclude that the four-dimensional domain wall solution is identical, upon up-lifting, to the 10-dimensional Calabi-Yau domain wall solution.

It is interesting to note that for the case of vanishing magnetic flux $\mu^a=0$, the above equations reduce to
\begin{equation}
 a\sim b^i\sim c^a\sim \mbox{const}\;, \quad \epsilon_ac^a=0\;,
\end{equation}
for the real parts and
\begin{equation} 
   \phi'=-\frac{1}{2}\frac{\tilde{\mathcal{K}}'}{\tilde{\mathcal{K}}}\; ,\quad
   \tilde{\mathcal{K}}_{ab}\partial_yw^a={\sqrt{\frac{\tilde{\mathcal{K}}}{\mathcal{K}}}}\epsilon_a\; ,\quad
   \partial_yv^i=-2\phi'v^i\; ,
\end{equation}
for the imaginary parts. These equations are mirror-symmetric to~\eqref{dweq} under the following correspondence
\begin{equation} 
	v^i\longleftrightarrow w^a \;,\quad\mathcal{K}_{ijk}\longleftrightarrow\tilde{\mathcal{K}}_{abc} \;,\quad e_i\longleftrightarrow \epsilon_a
\end{equation}
and can, therefore, be integrated in the same way. This fact is not surprising and reflects the original construction of half-flat mirror manifolds~\cite{Gurrieri:2002wz} as type II mirror duals of Calabi-Yau manifolds with electric NS flux. In the present context it suggests a mirror symmetry between heterotic Calabi-Yau compactifications with electric NS flux and heterotic compactifications based on the associated half-flat mirror manifold. 

\section{Conclusion}\label{Conclusion}

In this paper, we have studied 10-dimensional solutions of the heterotic string which involve a warped product of a four-dimensional domain wall with a six-dimensional internal space. Such solutions provide the general setting for heterotic compactifications with flux and on manifolds with $SU(3)$ structure. 

For the special case with vanishing flux and a constant dilaton the solution is a direct product of the $2+1$-dimensional domain wall world volume and a seven-dimensional manifold with $G_2$ holonomy. This $G_2$ manifold in turn consists of a six-dimensional half-flat manifold varying along the direction, $y$, transverse to the domain wall as specified by Hitchin's flow equations. We have shown that these 10-dimensional solutions form the basis for compactification on half-flat mirror manifolds without flux as carried out in Ref.~\cite{Gurrieri:2004dt}. Specifically, we have verified that the BPS domain walls in the four-dimensional $N=1$ supergravity theories associated to these compactifications precisely lift to our 10-dimensional solutions. 

We have further generalized this picture to include non-vanishing flux and a non-constant dilaton. In this case, the 10-dimensional space is still a direct product between the $2+1$-dimensional domain wall word volume and a seven-dimensional space. However, this seven-dimensional space now has $G_2$ structure rather than $G_2$ holonomy. It can also be thought of as the variation of a six-dimensional manifold along the direction $y$, where the variation is described by a generalized version of Hitchin's flow equations. The torsion classes of the allowed spaces are constrained by the relations~\eqref{tc}. In particular, these constraints imply that the six-dimensional manifolds are generalized half-flat and almost complex but, in general, no longer complex. Compared to Strominger's original class of complex, non-Kahler manifolds this opens up many more possibilities. In particular, flux compactifications on half-flat mirror manifolds are based on these solutions. These manifolds constitute a very large set: one such manifold is obtained for each Calabi-Yau three-fold (with a mirror) and a choice of electric NS flux on this three-fold~\cite{Gurrieri:2002wz}. 

Finally, we have obtained a class of solutions consisting of an exact Calabi-Yau three-fold with NS flux, which varies in its moduli space as one moves along the direction $y$. For the case of purely electric NS-NS flux, they are the natural candidate mirrors for the solutions based on $G_2$ holonomy manifolds. This is analogous to the original type II mirror symmetry correspondence with NS flux~\cite{Gurrieri:2002wz}. 

Our results open up new possibilities for heterotic string model building and they put heterotic half-flat compactifications on a more solid theoretical basis. It would be interesting to study the lift of these solutions to heterotic M-theory~\cite{Horava:1996ma,Witten:1996mz,Lukas:1998yy}. We also hope to follow up this work by studying heterotic models based on explicit classes of half-flat manifolds. 

\section*{Acknowledgements}
We would like to thank James Gray and Eran Palti for useful discussions. C.~M.~ is supported by a Berrow Foundation scholarship in association with Lincoln College Oxford.

\appendix
\section*{Appendix}

\section{Conventions}\label{Conventions}\setall

In this Appendix, we would like to summarize the conventions used throughout the paper. We begin with our index conventions. Ten-dimensional space-time $M_{10}$ is decomposed as $M_{10}=M_3\times\{y\}\times\hat{X}$ into three-dimensional Minkowski space $M_3$, a six-dimensional internal space $\hat{X}$ and a special direction $y$. We will also need to refer to the four-dimensional space $M_3\times\{y\}$ and the seven-dimensional space $Y=\{y\}\times \hat{X}$. The various index choices for these spaces are summarized in the table below.
\begin{center}
\begin{tabular}{|r|c|l|}
	\hline
	$10d$ & $M,N,...=0,1,...,9$\\
  \hline
  $7d$ & $m,n,p,...=3,4,...,9$\\
  \hline
  $6d$ & $u,v,...=4,5,...,9$\\
  \hline
  $4d$ & $\mu,\nu,...=0,1,2,3$\\
  \hline
  $3d$ & $\alpha,\beta,\gamma,...=0,1,2$\\
  \hline
  $1d$ & $M=\mu=m=3$\\
  \hline
\end{tabular}
\end{center}
These are curved indices and their tangent space counterparts will be underlined.\\

The 10-dimensional gamma matrices $\Gamma^{\underline{M}}$ are $32\times 32$ matrices which we choose to be purely imaginary. They satisfy the usual commutation relations
\begin{equation}
 \left\{\Gamma^{\underline{M}},\Gamma^{\underline{N}}\right\}=2\eta^{\underline{M}\underline{N}}\cdot\mathds{1}_{32}\; .
\end{equation}
The chirality operator in this basis is given by
\begin{equation}
	\Gamma^{\underline{11}}=\Gamma^{\underline{0}}\Gamma^{\underline{1}}...\Gamma^{\underline{9}}\; .
\end{equation}
Given the above choice of gamma matrices, the Majorana condition on 10-dimensional Dirac spinors $\epsilon$ is simply the reality condition $\epsilon=\epsilon^*$ and chiral spinors satisfy $\Gamma^{\underline{11}}\epsilon=\pm\epsilon$.

We choose the three-dimensional gamma matrices $\tilde{\gamma}^{\underline{\alpha}}$ to be real and the seven-dimensional gamma matrices $\gamma^{\underline{m}}$ to be purely imaginary. They satisfy
\begin{equation}
\{\tilde{\gamma}^{\underline{\alpha}},\tilde{\gamma}^{\underline{\beta}}\}=2\eta^{\alpha\beta}\mathds{1}_{2}\; ,\quad
\{\gamma^{\underline{m}},\gamma^{\underline{n}}\}=2\delta^{mn}\mathds{1}_{8}\; .
\end{equation}
Three-dimensional Majorana spinors $\rho$ and seven-dimensional Majorana spinor $\eta$ are then simply real spinors. As usual, anti-symmetrisation of gamma matrices is with strength one and is denoted by
\begin{equation}
\gamma^{\underline{m}_1...\underline{m}_k}=\gamma^{[\underline{m}_1}\gamma^{\underline{m}_2}...\gamma^{\underline{m}_k]}\; .
\end{equation}

With the above conventions we can decompose the 10-dimensional gamma matrices as
\begin{equation}
	\Gamma^{\underline{a}}=\tilde\gamma^{\underline{a}}\otimes\mathds{1}_8\otimes \sigma^{\underline{2}}\; ,\quad
	\Gamma^{\underline{m}}=\mathds{1}_2\otimes\gamma^{\underline{m}}\otimes \sigma^{\underline{1}} \; ,
\end{equation}
where $\sigma^{\underline{\mu}}$ denote the usual Pauli matrices together with $\sigma^{\underline{0}}=\mathds{1}_2$. A 10-dimensional Majorana-Weyl spinor $\epsilon$ can be constructed from three- and seven-dimensional Majorana spinors $\rho$ and $\eta$ by writing
\begin{equation}
	\epsilon=\rho\otimes\eta\otimes\theta,
\end{equation}
where $\theta$ is an eigenvector of the third Pauli matrix $\sigma^{\underline{3}}$ whose eigenvalue determines the chirality of $\epsilon$. It will also be useful to express the seven-dimensional Majorana spinor as
\begin{equation}
	\eta=\frac{1}{\sqrt{2}}(\eta_++\eta_-)\; , \label{etadecompA}
\end{equation}
where $\eta_\pm$ are six-dimensional chiral spinors satisfying $\gamma^{\underline{3}}\eta_\pm=\pm\eta_\pm$ and $\eta_-=\eta_+^*$.

\section{Torsion classes}\label{Torsion classes} \setall

In this Appendix, we review some facts on $G$-structures, in particular $SU(3)$ and $G_2$ structures, which will be used in the main text. We will be brief and refer to the literature~\cite{ChiossiSalamon,Grana:2005jc, LopesCardoso:2002hd, Kaste:2003zd} for a more detailed discussion.

For an $n$-dimensional manifold, the structure group of its frame bundle is in general contained in $GL(n,R)$. The manifold is said to admit a $G$-structure, where $G\subset GL(n,R)$ is a sub-group, if a sub-bundle of the frame bundle with structure group $G$ exists. Alternatively, a $G$-structure can also be characterized by globally defined spinors on the manifold or a set of globally defined forms. In the present paper, we are interested in $G_2$ structures on seven-dimensional manifolds and $SU(3)$ structures on six- and seven-dimensional manifolds. 
The invariant forms which characterize these various structures are summarized in the following table.
\begin{center}
\begin{tabular}{|c|c|l|}
	\hline
  Dimension & Group $G$ & Tensors\\
  \hline
  $7d$ & $G_2$ & $\varphi$, $\Phi$\\
  \hline
  $7d$ & $SU(3)$ & $J$, $\Omega$, $\alpha$\\
  \hline
  $6d$ & $SU(3)$ & $J$, $\Omega$\\
  \hline
\end{tabular}
\end{center}
For a $G$-structure there exists a connection $\nabla^{(T)}$, in general with torsion, satisfying $\rm{hol}(\hat\nabla)\subset G$. The tensors characterizing the $G$-structure are covariantly constant with respect to this connection. The con-torsion $\kappa$ contained in $\nabla^{(T)}$ is can be viewed as a one-form taking values in the Lie algebra of $so(n)$ and can be decomposed as
\begin{equation}
	\kappa_m=\kappa_m^0+\kappa_m^G\; .
\end{equation}
Here $\kappa^G$ takes values in ${\cal L}(G)$, the Lie algebra of $G$, and $\kappa^0$ in its orthogonal complement ${\cal L}(G)^\bot$ in $so(n)$. The reason for this decomposition is that the action of $\kappa_m^G$ on the $G$-invariant tensors vanishes. Hence, the fact that the invariant tensors are covariantly constant under $\nabla^{(T)}$ and that  the holonomy of $\nabla^{(T)}$ is contained in $G$ only depends on $\kappa_m^0$. For this reason, $\kappa_m^0$ is also called the intrinsic (con)-torsion. It can be decomposed into its irreducible representation content under the group $G$. These irreducible parts of $\kappa_m^0$ are called torsion classes and they can be used to characterize the $G$-structure.\\

We begin reviewing this more concretely for $G_2$ structures on a seven-dimensional manifold. The torsion is a one form, with its one-form index transforming as a fundamental of $SO(7)$,  and otherwise taking values in the adjoint of $SO(7)$. Hence, the two relevant decompositions under $G_2$ are
\begin{equation}
{\bf 7}_{SO(7)}\rightarrow{\bf 7}_{G_2}\; ,\quad{\bf 21}_{SO(7)}\rightarrow ({\bf 7}+{\bf 14})_{G_2}\; .
\end{equation}
The intrinsic torsion only takes values in ${\cal L}(G_2)^\bot={\bf 7}_{G_2}$ and, hence, its $G_2$ representation content is given by
\begin{equation}
 {\bf 7}\otimes{\bf 7}={\bf 1}+{\bf 14}+{\bf 27}+{\bf 7}\; .
\end{equation} 
The representations on the right-hand side correspond to the four torsion classes ${\cal X}_1,\ldots,{\cal X}_4$ associated to a $G_2$ structure and, hence, the con-torsion takes values
\begin{equation}
	\kappa^0\in {\cal X}_1\oplus {\cal X}_2\oplus {\cal X}_3\oplus {\cal X}_4\; .
\end{equation}
The $G_2$ structure can be characterized by a seven-dimensional Majorana spinor $\eta$ or, alternatively, by a three-form $\varphi$ and four-form $\Phi$. In terms of the spinor $\eta$, these forms can be written as
\begin{equation}
	\varphi_{mnp}=-i \eta^\dagger\gamma_{mnp}\eta\; ,\quad
\Phi_{mnpq}=\eta^\dagger\gamma_{mnpq}\eta\; . \label{sping2}
\end{equation}
It is easy to verify that
\begin{equation}
	\varphi=*_7\Phi\; ,
\end{equation}
where $*_7$ is the seven-dimensional Hodge star with respect to the metric induced by the $G_2$ structure.
A straightforward computation shows that the exterior derivatives of these forms depend on the torsion classes and are given by
\begin{equation}
	d_7\varphi=4{\cal X}_1\Phi+3{\cal X}_4\wedge\varphi-*_7{\cal X}_3\; ,\quad
	d_7\Phi=4{\cal X}_4\wedge\Phi-2*_7{\cal X}_2\; , \label{g2torsion}
\end{equation}
where $*_7$ is the seven-dimensional exterior derivative. Often these equations offer the most straightforward way to determine the torsion classes by computing the exterior derivatives of $\varphi$ and $\Phi$.\\

We now move on to $SU(3)$ structures on six-dimensional manifolds. The torsion takes values in the Lie-algebra $so(6)$ while its one-form index transforms under the fundamental of $SO(6)$. Hence, the relevant $SU(3)$ decompositions read
\begin{equation}
 {\bf 6}_{SO(6)}\rightarrow ({\bf 3}+\bar{\bf 3})_{SU(3)}\; ,\quad
 {\bf 15}_{SO(6)}\rightarrow({\bf 1}+{\bf 3}+\bar{\bf 3}+{\bf 8})_{SU(3)}\; .
\end{equation} 
Since, ${\cal L}(SU(3))^\bot={\bf 1}+{\bf 3}+\bar{\bf 3}$, the intrinsic torsion contains the irreducible $SU(3)$ representations
\begin{equation}
	({\bf 3}+\bar{\bf 3})\otimes({\bf 1}+{\bf 3}+\bar{\bf 3})=({\bf 1}+{\bf 1})+({\bf 8}+{\bf 8})+
	                                          ({\bf 6}+\bar{\bf 6})+({\bf 3}+\bar{\bf 3})+({\bf 3}+\bar{\bf 3})\; ,
\end{equation}
which gives rise to the five torsion classes
\begin{equation}\label{Ws}
	\kappa^0\in W_1\oplus W_2\oplus W_3\oplus W_4\oplus W_5\; .
\end{equation}
Properties of the six-dimensional manifold can be characterized by these five torsion classes as indicated in the table below.
\begin{center}
\begin{tabular}{|c|c|}
	\hline
  Vanishing Torsion Classes & Properties (Name)\\
  \hline
  $W_1=W_2=0$ & Complex\\
  \hline
  $W_1=W_3=W_4=0$ & Symplectic\\
  \hline
  $W_1=W_2=W_3=W_4=0$ & Kahler\\
  \hline
  $W_{1-}=W_{2-}=W_4=W_5=0$ & Half-flat\\
  \hline
	$W_1=W_2=W_3=W_4=W_5=0$ & Calabi-Yau\\
  \hline
\end{tabular}
\end{center}
An $SU(3)$ structure is determined by a six-dimensional Weyl spinor $\eta_+$ and its conjugate $\eta_-=\eta_+^*$ or, alternatively, by a two-form $J$ and a three-form $\Omega$. In terms of the spinors, these forms can be written as
\begin{equation}
	J_{uv}=- i\eta_+^\dagger\gamma_{uv}\eta_+\; ,\quad
	\Omega_{uvw}=\eta^\dagger_+\gamma_{uvw}\eta_-\; . \label{spinsu3}
\end{equation}
The exterior derivatives of $J$ and $\Omega$ are explicitly given by
 \begin{equation}\label{su3torsion}
	dJ=-\frac{3}{2}{\rm Im}(W_1\bar\Omega)+W_4\wedge J+W_3\; ,\qquad
	d\Omega=W_1J\wedge J+W_2\wedge J+\bar W_5\wedge\Omega\; .
\end{equation}
As a consequence of their $SU(3)$ transformation properties the torsion classes satisfy the following useful constraints
\begin{equation}
	W_3\wedge J=W_3\wedge\Omega=W_2\wedge J\wedge J=0\; .
\end{equation}
An alternative way to define an $SU(3)$ structure is to start with the pair of forms $(J,\Omega)$ and require the conditions
\begin{equation}
	J\wedge J\wedge J=\frac{3}{4}i\Omega\wedge\bar\Omega, \quad \Omega\wedge J=0\; . \label{su3def}
\end{equation}
which will be used in the main part of the text.\\

An $SU(3)$ structure on a seven-dimensional manifold can be defined by a triplet $(J,\Omega,\alpha)$ of forms, where $J$ is a two-form and $\Omega$ a three-form, as before, and $\alpha$ is a one-form. Intuitively, $\alpha$ singles out a special direction and a complementary six-dimensional space on which $J$ and $\Omega$ can be thought of as defining an $SU(3)$ structure in the six-dimensional sense. In addition to the usual conditions~\eqref{su3def} for a six-dimensional $SU(3)$ structure, its seven-dimensional counterpart must satisfy a number of additional relations which involve $\alpha$. We will not give these relations explicitly but instead refer to Refs.~\cite{Lukas:2004ip}. From the spinor expressions~\eqref{sping2} and \eqref{spinsu3} together with Eq.~\eqref{etadecompA} one can show that a seven-dimensional $SU(3)$ structure gives rise to a $G_2$ structure via
\begin{equation}
\varphi=\alpha\wedge J+\Omega_-\; ,\quad \Phi=\alpha\wedge \Omega_++\frac{1}{2}J\wedge J\; .
\end{equation}

\section{Moduli space geometry}\label{Moduli space geometry}\setall

In this Appendix, we collect some information about the moduli space geometry of Calabi-Yau manifolds which will be used throughout the paper. This material is well-known~\cite{Candelas:1990pi} and is merely included for our and the reader's convenience.\\

We consider a Calabi-Yau three-fold $X$ with Kahler form $J$ and holomorphic $(3,0)$-form $\Omega$ and Hodge numbers $h^{1,1}(X)$ and $h^{2,1}(X)$. We have a basis $\{\omega_i\}$, where $i,j,\ldots =1,\ldots ,h^{1,1}(X)$ of harmonic two-forms and a symplectic basis $\{\alpha_A,\beta^A\}$, where $A,B,\ldots = 0,1,\ldots ,h^{2,1}(X)$, of harmonic three forms. We can define the moduli by expanding $J$ and $\Omega$ in terms of these basis forms as
\begin{equation}
 J=v^i\omega_i\; ,\quad \Omega={\cal Z}^A\alpha_A-\mathcal{G}_A\beta^A\; .
\end{equation} 
Here, $v^i$ are the Kahler moduli and we denote their complexification by $T^i=b^i+iv^i$. The ${\cal Z}^A$ are the projective complex structure moduli. Their affine counterparts are defined by $Z^a={\cal Z}^a/{\cal Z}^0$, where $a,b,\ldots = 1,\ldots ,h^{2,1}(X)$, and split up into real and imaginary parts as $Z^a=c^a+iw^a$. The harmonic $(2,1)$ and $(1,2)$ forms associated to $Z^a$ and $\bar{Z}^a$ are denoted by $\chi_a$ and $\bar{\chi}_{a}$, respectively. We also introduce the triple intersection numbers
\begin{equation}
 {\cal K}_{ijk}=\int_X\omega_i\wedge\omega_j\wedge\omega_k
\end{equation}
of $X$ as well as the triple intersection numbers $\tilde{\cal K}_{abc}$ of the mirror $\tilde{X}$ of $X$.\\

After this set-up we begin with the Kahler moduli space. The moduli space metric in the large radius limit can be written as
\begin{equation}
K^{(1)}_ {ij}=\frac{1}{4\mathcal{V}}\int_X\omega_i\wedge*\omega_j\; ,
\end{equation}
where ${\cal V}$ is the volume of $X$. In order to describe this metric explicitly it is useful to introduce the following functions
\begin{equation}
 {\cal K}={\cal K}_{ijk}v^iv^jv^k\; ,\quad {\cal K}_i={\cal K}_{ijk}v^jv^k\; ,\quad {\cal K}_{ij}={\cal K}_{ijk}v^k\; \label{metric1}
\end{equation} 
 of the Kahler moduli $v^i$. It is easy to verify that the volume can be written as
\begin{equation}
 6{\cal V}={\cal K}=\int_XJ\wedge J\wedge J\; .
\end{equation} 
It can then be shown that the metric~\eqref{metric1}, as a function of the complexified fields $T^i$, is Kahler and can be obtained as
\begin{equation}
 K^{(1)}_{ij}=\frac{\partial^2K^{(1)}}{\partial T^i\partial\bar{T}^j}\; ,\quad K^{(1)}=-\ln\left(\frac{4}{3}{\cal K}\right)\; , \label{K1}
\end{equation} 
where $K^{(1)}$ is the Kahler potential. Explicitly, this means
\begin{equation}
 K^{(1)}_i\equiv \frac{\partial K^{(1)}}{\partial T^i}=\frac{3i}{2}\frac{\mathcal{K}_i}{\mathcal{K}}\; ,\quad
 K^{(1)}_ {ij}=\frac{9}{4}\frac{\mathcal{K}_i\mathcal{K}_j}{\mathcal{K}^2}-\frac{3}{2}\frac{\mathcal{K}_{ij}}{\mathcal{K}}\; .
\end{equation}

Now we move on to the complex structure moduli space. As before, the moduli space metric is Kahler and takes form
\begin{equation}\label{CSmetric}
	K^{(2)}_{a b}=-\frac{\int_X\chi_a\wedge\bar\chi_b}{\int_X\Omega\wedge\bar\Omega}
	=\frac{\partial^2 K^{(2)}}{\partial Z^a\partial \bar{Z}^b}\; .
\end{equation}
with the Kahler potential
\begin{equation}\label{CSKpot}
	K^{(2)}=-\ln\left(i\int_X \Omega\wedge\bar\Omega\right)=-\ln\left(i(\bar Z^A\mathcal{G}_A-Z^A\bar{\mathcal{G}}_A)\right)\; . 
\end{equation}
Here, ${\cal G}$ is the pre-potential, a holomorphic function of the fields ${\cal Z}^A$ which is homogeneous of degree two and ${\cal G}_A=\partial{\cal G}/\partial{\cal Z}^A$. That this Kahler potential does indeed lead to the correction metric~\eqref{CSmetric} can be verified by using Kodaira's formula
\begin{equation}\label{Kodaira}
	\frac{\partial\Omega}{\partial z^a}=-\frac{\partial K^{(2)}}{\partial Z^a}\Omega+\chi_a\; .
\end{equation}
In terms of $\Omega$ the volume of the Calabi-Yau manifold can be expressed as
\begin{equation}\label{OmegaandV}
	\mathcal{V}=\frac{i}{\|\Omega\|^2}\int_X \Omega\wedge \bar\Omega\; , 
\end{equation}
where $3!\|\Omega\|^2=\Omega_{uvw}\bar\Omega^{uvw}$.

In the large complex structure limit, the pre-potential has the simple form
\begin{equation}\label{prepotential}
	\mathcal{G}=-\frac{1}{6}\tilde{\mathcal{K}}_{abc}\frac{Z^aZ^bZ^c}{Z^0}\; .
\end{equation}
In this case, we can introduce the functions
\begin{equation}
	\tilde{\mathcal{K}}=\tilde{\mathcal{K}}_{abc}w^aw^bw^c\;, \quad \tilde{\mathcal{K}}_a=\tilde{\mathcal{K}}_{abc}w^bw^c\; ,\quad
	\tilde{\mathcal{K}}_{ab}=\tilde{\mathcal{K}}_{abc}w^c\; ,
\end{equation}
and write the complex structure moduli space metric as
\begin{equation}
K^{(2)}_{a}=\frac{\partial K^{(2)}}{\partial Z^a}=\frac{3i}{2}\frac{\tilde{\mathcal{K}}_a}{\tilde{\mathcal{K}}}\; ,\quad
K^{(2)}_{a\bar b}=\frac{9}{4}\frac{\tilde{\mathcal{K}}_a\tilde{\mathcal{K}}_b}{\tilde{\mathcal{K}}^2}-\frac{3}{2}\frac{\tilde{\mathcal{K}}_{ab}}{\tilde{\mathcal{K}}}\; ,
\end{equation}
in complete analogy with the equations in the Kahler moduli sector.\\

The total Kahler potential $K$ can now be written as $K=K^{(1)}+K^{(2)}+K^{(S)}$ where 
\begin{equation}
	K^{(S)}=-\ln\left(i(\bar S-S)\right)=-\ln\left(2e^{-2\phi}\right)\; ,
\end{equation}
is the contribution from the dilaton $S=a+ie^{-2\phi}$. From Eqs.~\eqref{K1}, \eqref{CSKpot} and \eqref{OmegaandV} this leads to the useful formula
\begin{equation}\label{eK2}
	e^{K/2}=\frac{e^\phi}{4\mathcal{V}\|\Omega\|}\; .
\end{equation}



\begin{thebibliography}{99}

\bibitem{Candelas:1985en}
  P.~Candelas, G.~T.~Horowitz, A.~Strominger and E.~Witten,
  ``Vacuum Configurations For Superstrings,''
  Nucl.\ Phys.\ B {\bf 258}, 46 (1985).
  
\bibitem{Braun:2005ux}
  V.~Braun, Y.~H.~He, B.~A.~Ovrut and T.~Pantev,
  ``A heterotic standard model,''
  Phys.\ Lett.\ B {\bf 618}, 252 (2005)
  [arXiv:hep-th/0501070].
  V.~Braun, Y.~H.~He, B.~A.~Ovrut and T.~Pantev,
  ``A standard model from the E(8) x E(8) heterotic superstring,''
  JHEP {\bf 0506}, 039 (2005)
  [arXiv:hep-th/0502155].  
  
\bibitem{Bouchard:2005ag}
  V.~Bouchard and R.~Donagi,
  ``An SU(5) heterotic standard model,''
  Phys.\ Lett.\ B {\bf 633}, 783 (2006)
  [arXiv:hep-th/0512149].  
  
\bibitem{Anderson:2009mh}
  L.~B.~Anderson, J.~Gray, Y.~H.~He and A.~Lukas,
  ``Exploring Positive Monad Bundles And A New Heterotic Standard Model,''
  JHEP {\bf 1002} (2010) 054
  [arXiv:0911.1569 [hep-th]].
  
\bibitem{Giddings:2001yu}
  S.~B.~Giddings, S.~Kachru and J.~Polchinski,
  ``Hierarchies from fluxes in string compactifications,''
  Phys.\ Rev.\  D {\bf 66} (2002) 106006
  [arXiv:hep-th/0105097].
  
\bibitem{Strominger:1986uh}
  A.~Strominger,
  ``Superstrings with Torsion,''
  Nucl.\ Phys.\  B {\bf 274} (1986) 253.
  
\bibitem{Gurrieri:2004dt}
  S.~Gurrieri, A.~Lukas and A.~Micu,
  ``Heterotic on half-flat,''
  Phys.\ Rev.\  D {\bf 70} (2004) 126009
  [arXiv:hep-th/0408121].
  
\bibitem{Gurrieri:2007jg}
  S.~Gurrieri, A.~Lukas and A.~Micu,
  ``Heterotic String Compactifications on Half-flat Manifolds II,''
  JHEP {\bf 0712} (2007) 081
  [arXiv:0709.1932 [hep-th]].
  
\bibitem{deCarlos:2005kh}
  B.~de Carlos, S.~Gurrieri, A.~Lukas and A.~Micu,
  ``Moduli stabilisation in heterotic string compactifications,''
  JHEP {\bf 0603} (2006) 005
  [arXiv:hep-th/0507173].
  
\bibitem{Gurrieri:2005af}
  S.~Gurrieri,
  ``Compactifications On Half-Flat Manifolds,''
  Fortsch.\ Phys.\  {\bf 53} (2005) 278.
 
\bibitem{Hitchin}
 N.~Hitchin,
  ``Stable forms and special metrics",
  ``Proceedings of the Congress in memory of Alfred Gray", (eds M. Fernandez and J. Wolf), AMS Contemporary Mathematics Series,
  arXiv:math/0107101v1 [math.DG].
 
\bibitem{ChiossiSalamon}
  S.~Chiossi and S.~Salamon,
  ``The intrinsic torsion of SU(3) and G2 structures,''
  [arXiv:0202282v1 [math.DG]]. 
  
\bibitem{Grana:2005jc}
  M.~Grana,
  ``Flux compactifications in string theory: A comprehensive review,''
  Phys.\ Rept.\  {\bf 423} (2006) 91
  [arXiv:hep-th/0509003].
  
\bibitem{LopesCardoso:2002hd}
  G.~Lopes Cardoso, G.~Curio, G.~Dall'Agata, D.~Lust, P.~Manousselis and G.~Zoupanos,
  ``Non-Kaehler string backgrounds and their five torsion classes,''
  Nucl.\ Phys.\  B {\bf 652} (2003) 5
  [arXiv:hep-th/0211118].
  
\bibitem{Gran:2005wf}
  U.~Gran, P.~Lohrmann and G.~Papadopoulos,
  ``The spinorial geometry of supersymmetric heterotic string backgrounds,''
  JHEP {\bf 0602} (2006) 063
  [arXiv:hep-th/0510176].
    
\bibitem{Green:1987mn}
  M.~B.~Green, J.~H.~Schwarz and E.~Witten,
  ``Superstring Theory. Vol. 2: Loop Amplitudes, Anomalies And Phenomenology,''
{\it  Cambridge, Uk: Univ. Pr. ( 1987) 596 P. ( Cambridge Monographs On Mathematical Physics)}

\bibitem{polchinski}
 J.~Polchinski, ``String Theory", Vol.~2, Cambridge Monographs on Mathematical Physics, 1998.  
 
\bibitem{Benmachiche:2008ma}
  I.~Benmachiche, J.~Louis and D.~Martinez-Pedrera,
  ``The effective action of the heterotic string compactified on manifolds with
  SU(3) structure,''
  Class.\ Quant.\ Grav.\  {\bf 25} (2008) 135006
  [arXiv:0802.0410 [hep-th]].

\bibitem{Ali:2006gd}
  T.~Ali and G.~B.~Cleaver,
  ``The Ricci Curvature of Half-flat Manifolds,''
  JHEP {\bf 0705} (2007) 009
  [arXiv:hep-th/0612171].
  
\bibitem{Ali:2007ra}
  T.~Ali and G.~B.~Cleaver,
  ``A Note on the Standard Embedding on Half-Flat Manifolds,''
  JHEP {\bf 0807} (2008) 121
  [arXiv:0711.3248 [hep-th]].
  
\bibitem{Gurrieri:2002wz}
  S.~Gurrieri, J.~Louis, A.~Micu and D.~Waldram,
  ``Mirror symmetry in generalized Calabi-Yau compactifications,''
  Nucl.\ Phys.\  B {\bf 654} (2003) 61
  [arXiv:hep-th/0211102].

\bibitem{GVW}
S.~Gukov, C.~Vafa and E.~Witten,
``CFT's from Calabi-Yau four-folds,''
Nucl.\ Phys.\ B {\bf 584} (2000) 69
[Erratum-ibid.\ B {\bf 608} (2001) 477]
[arXiv:hep-th/9906070].
  
\bibitem{Eto:2003bn}
  M.~Eto and N.~Sakai,
  ``Solvable models of domain walls in N = 1 supergravity,''
  Phys.\ Rev.\  D {\bf 68} (2003) 125001
  [arXiv:hep-th/0307276].
  
\bibitem{Cvetic:1996vr}
  M.~Cvetic and H.~H.~Soleng,
  ``Supergravity domain walls,''
  Phys.\ Rept.\  {\bf 282} (1997) 159
  [arXiv:hep-th/9604090].

\bibitem{Wess:1992cp}
  J.~Wess and J.~Bagger,
  ``Supersymmetry and supergravity,''
{\it  Princeton, USA: Univ. Pr. (1992) 259 p}  
    
\bibitem{Mayer:2004sd}
  C.~Mayer and T.~Mohaupt,
  ``Domain Walls, Hitchin's Flow Equations and $G_2$-Manifolds,''
  Class.\ Quant.\ Grav.\  {\bf 22} (2005) 379
  [arXiv:hep-th/0407198].
  
\bibitem{Smyth:2009fu}
  P.~Smyth and S.~Vaula,
  ``Domain wall flow equations and SU(3)xSU(3) structure compactifications,''
  Nucl.\ Phys.\  B {\bf 828} (2010) 102
  [arXiv:0905.1334 [hep-th]].
  
\bibitem{Gauntlett:2002sc}
  J.~P.~Gauntlett, D.~Martelli, S.~Pakis and D.~Waldram,
  ``G-structures and wrapped NS5-branes,''
  Commun.\ Math.\ Phys.\  {\bf 247} (2004) 421
  [arXiv:hep-th/0205050].
  
\bibitem{Friedrich}
  T.~Friedrich and S.~Ivanov,
  ``Killing spinor equations in dimension 7 and geometry of integrable $G_{2}$-manifolds,''
  Journal of Geometry and Physics\ {\bf 48} (2003) 1-11.
  [arXiv:math/0112201].
  
\bibitem{Held:2010az}
  J.~Held, D.~Lust, F.~Marchesano and L.~Martucci,
  ``DWSB in heterotic flux compactifications,''
  arXiv:1004.0867 [hep-th].

\bibitem{Horava:1996ma}
  P.~Horava and E.~Witten,
  ``Eleven-Dimensional Supergravity on a Manifold with Boundary,''
  Nucl.\ Phys.\  B {\bf 475} (1996) 94
  [arXiv:hep-th/9603142].
  
\bibitem{Witten:1996mz}
  E.~Witten,
  ``Strong Coupling Expansion Of Calabi-Yau Compactification,''
  Nucl.\ Phys.\  B {\bf 471}, 135 (1996)
  [arXiv:hep-th/9602070].
  
\bibitem{Lukas:1998yy}
  A.~Lukas, B.~A.~Ovrut, K.~S.~Stelle and D.~Waldram,
  ``The universe as a domain wall,''
  Phys.\ Rev.\  D {\bf 59}, 086001 (1999)
  [arXiv:hep-th/9803235].

\bibitem{Kaste:2003zd}
  P.~Kaste, R.~Minasian and A.~Tomasiello,
  ``Supersymmetric M-theory compactifications with fluxes on  seven-manifolds
  and G-structures,''
  JHEP {\bf 0307} (2003) 004
  [arXiv:hep-th/0303127].
  
\bibitem{Lukas:2004ip}
  A.~Lukas and P.~M.~Saffin,
  ``M-theory compactification, fluxes and AdS(4),''
  Phys.\ Rev.\  D {\bf 71} (2005) 046005
  [arXiv:hep-th/0403235].

\bibitem{Candelas:1990pi}
  P.~Candelas and X.~de la Ossa,
  ``Moduli space of Calabi-Yau manifolds,"
  Nucl.\ Phys.\  B {\bf 355} (1991) 455.

\end{thebibliography}
\end{document}